\documentclass[aps,pra,preprint]{revtex4}

\usepackage{epsfig}
\usepackage{subfigure}

\begin{document}

\title{Comparison Tests of Variable-Stepsize Algorithms for Stochastic Ordinary Differential Equations of Finance}
\author{Yin Mei Wong$^{1}$ and Joshua Wilkie$^{2}$}
\affiliation{
$^{1}$ Innovative Stochastic Algorithms, Vancouver, British Columbia V6E 2C8, Canada\\
$^{2}$ Department of Chemistry, Simon Fraser University, Burnaby, British Columbia V5A 1S6, Canada}

\date{\today}
 
\begin{abstract}
Since the introduction of the Black-Scholes model stochastic processes have played an increasingly important role
in mathematical finance. In many cases prices, volatility and other quantities can be modeled using stochastic
ordinary differential equations. Available methods for solving such equations have until recently been markedly inferior to
analogous methods for deterministic ordinary differential equations. Recently, a number of methods which employ
variable stepsizes to control local error have been developed which appear to offer greatly improved speed and accuracy.
Here we conduct a comparative study of the performance of these algorithms for problems taken from the 
mathematical finance literature.

\end{abstract}

\maketitle

\section{Introduction}

Stochastic processes play an increasingly important role in mathematical finance as evidenced by the large 
and growing literature on stochastic volatility models\cite{Pearson,CIR,Cox,Davis,HW,Hull,Scott,Nel,AB,DGH,DK,Chern}. Often these theories are expressed in
terms of stochastic ordinary differential equations (SODEs). Examples include the Cox-Ingersoll-Ross\cite{Pearson,CIR,Cox}, Hull-White\cite{Davis,HW,Hull}, 
Log Ornstein-Uhlenbeck\cite{Scott}, Nelson\cite{Nel}, Affine\cite{DK} and Log-linear\cite{Chern} models of stochastic volatility. Other schemes like ARCH models\cite{Nel,ARCH}
use discrete time difference equations which can be viewed as approximations to diffusions\cite{Nel}, and which are
often favored for computational and other reasons. SODE based models tend to have closer relationships to fundamental theory, but
have the drawback that analytic solutions are rarely known. In general these equations must be solved using numerical approximation schemes.

Numerical methods for SODEs have a long history\cite{KP} but until recently these algorithms have not
achieved the speed and accuracy characteristic of analogous methods for deterministic ordinary differential 
equations (ODEs)\cite{Hair}. This is partly due to the lack of variable-stepsize algorithms which allow for the control of
local error, and partly due to a lack of sufficiently high order algorithms. The MAPLE Stochastic Package\cite{MAP}, for example, fails to 
include variable-stepsize routines and most methods are of rather low order. Potential solutions to both of these problems have been reported in the last few years. Discussions of variable stepsize strategies for SODEs\cite{Gai,Lam} and some basic observations
regarding Taylor expansions for SODEs\cite{Wilk} have led to the emergence of
a number of published\cite{WC} and unpublished\cite{ANISE} variable-stepsize codes. These algorithms also
have a number of promising additional features such as linear scaling of computational cost with numbers of Wiener
processes. 

In this manuscript we perform a variety of tests to see whether the algorithms give the expected improved performance
and accuracy. We refer to the method developed by Wilkie and \c{C}etinba\c{s}\cite{Wilk,WC} as SDE9, and the unpublished commercial method\cite{ANISE} as ANISE. We will not attempt to discuss how these codes work but merely focus on their performance. We 
do not consider other variable-stepsize codes such as the weak method introduced in Ref. \cite{WC} since they have restricted 
domains of applicability.

The methods SDE9 and ANISE when applied to an It\^{o} stochastic differential equation 
\begin{equation}
dX_t=a(X_t,t)dt+\sum_{i=1}^m b_i(X_t,t) dW_{it}
\end{equation}
for an observable $X_t$ with Wiener processes $W_{it}$ require knowledge
of the partial derivatives of the solutions, i.e.,
\begin{eqnarray}
\frac{\partial X_t}{\partial W_{it}}&=&b_i(X_t,t)\\
\frac{\partial X_t}{\partial t}&=&a(X_t,t)-\frac{1}{2}\sum_{i=1}^m\frac{\partial b_i(X_t,t) }{\partial W_{it}}.
\end{eqnarray}
All of the problems we consider are formulated with It\^{o} stochastic differential equations and we provide these derivatives for each problem.
ANISE and SDE9 are more easily applied to Stratonovich stochastic differential equations
\begin{equation}
dX_t=a(X_t,t)dt+\sum_{i=1}^m b_i(X_t,t)\circ dW_{it}
\end{equation}
for which
\begin{eqnarray}
\frac{\partial X_t}{\partial W_{it}}&=&b_i(X_t,t)\\
\frac{\partial X_t}{\partial t}&=&a(X_t,t).
\end{eqnarray}
Extensions to jump diffusions\cite{Chern} are also straightforward, but are not considered here.

Our study shows that both SDE9 and ANISE yield accurate solutions to a wide variety of stochastic volatility models. ANISE tends to
be about twice as fast as SDE9. For one problem we find that ANISE performs hundreds of times faster than SDE9. Both methods provide
a means of obtaining high accuracy solutions to SODE problems, and may prove to be useful quantitative tools for further research
in mathematical finance.

In section II we explore Monte-Carlo convergence of numerically calculated means and variances of price and volatilities for seven
stochastic volatility models taken from the finance literature. Section III examines the accuracies of the algorithms for individual
trajectories.

\section{Monte-Carlo Convergence Tests}

Here our goal is to test the accuracy and compare computational performance for the ANISE and SDE9 numerical methods for SODEs discussed in the introduction. To do this we compare exact and numerically calculated average quantities like mean price and mean volatility for a selection of models from the finance literature.

For each model we compare known exact average quantities $\overline{X_t}$
to the numerical averages $\overline{X^{approx.}}=\frac{1}{N}\sum_{j=1}^N X_t^{(j)}$ computed from individual stochastic evolutions 
$X_t^{(j)}$ for $j=1,\dots, N$, obtained using the SODE methods. We examine convergence 
to the exact solution by varying the number of trajectories $N$. For each observable $\overline{X_t}$
we calculate the base ten log of the mean relative error, 
\begin{equation}
\log_{10}[\frac{|\overline{X_t}-\overline{X_t^{approx.}}|}{{\rm max}\{|\overline{X_t}|,|\overline{X_t^{approx.}}|\}}], 
\end{equation}
and plot this against time. This specific denominator, ${\rm max}\{|\overline{X_t}|,|\overline{X_t^{approx.}}|\}$, is chosen since some 
observables $X_t$ pass through zero and relative error can therefore blow up.
We also examine the relative CPU times for the two methods. All calculations were performed on a 600 MHz Alpha processor with a
requested tolerance of $10^{-12}$.

\subsection{Nelson Model}

In the Nelson model\cite{Nel} the log-price $P_t$ and volatility $V_t$ obey
\begin{eqnarray}
dP_t&=&\sqrt{V_t} dW_{1t}\nonumber \\
dV_t&=&\theta (\omega-V_t) dt+\sqrt{2\lambda\theta}V_tdW_{2t},\nonumber
\end{eqnarray}
where the normally distributed real stochastic differentials are uncorrelated and have $\overline{dW_{it}}=0$ and
$\overline{dW_{it}^2}=dt$. Thus, this model has two Wiener processes and two equations. 

The derivatives required by the SODE methods are given in Table \ref{table:D1}. We employed a time step $dt=.1$
and integrated to 100.
\begin{table}[h]
\begin{center}
\begin{tabular}{c|ccc}
~ $X_t$ ~ & ~ $\frac{\partial X_t}{\partial t}$ ~ & ~ $\frac{\partial X_t}{\partial W_{1t}}$ ~ & ~ $\frac{\partial X_t}{\partial W_{2t}}$ ~ \\ \hline 
 $P_t$ & 0 & $\sqrt{V_t}$ & 0 \\ $V_t$ & ~ ~$\theta(\omega-V_t)-\lambda\theta V_t$ ~~ & 0 & $\sqrt{2\lambda\theta}~V_t$ \\ 
\end{tabular}
\end{center}
\caption{ Derivatives for Nelson Model. }
\label{table:D1}
\end{table}
As with ODE methods, the intermediate steps taken in ANISE and SDE9 do not
necessarily reflect certain aspects of the true solutions such as the 
positivity of $V_t$. To avoid floating point problems one thus programs 
$\sqrt{|V_t|}$ rather than $\sqrt{V_t}$. The actual solutions returned for 
$V_t$ will of course satisfy positivity as we will see in section III where 
we explore the accuracy of individual trajectories.

We explored convergence for the average quantities $\overline{ P_t}$, ${\rm var}(P_t)$, $\overline{V_t}$, and ${\rm var}(V_t)$. The 
known exact solutions for these quantities are given by
\begin{eqnarray}
\overline{ P_t}&=&P_0\\
{\rm var}(P_t)&=&\omega t +\frac{\omega-V_0}{\theta}(e^{-\theta t}-1)\\
\overline{V_t}&=&V_0e^{-\theta t}+\omega(1-e^{-\theta t})\\
{\rm var}(V_t)&=&V_0^2e^{-2\theta(1-\lambda)t}+\frac{\omega^2}{1-\lambda}(1-e^{-2\theta(1-\lambda)t})\nonumber \\
&-&\frac{2\omega(\omega-V_0)}{1-2\lambda}(e^{-\theta t}-e^{-2\theta(1-\lambda )t})-\overline{V_{t}}^2.
\end{eqnarray}
We chose parameters $\theta=.035$, $\omega=.636$, $\lambda=.296$, and set initial price and volatility to $P_0=.5$ and $V_0=.029$. 

\begin{figure}[h]
  \begin{center}
    \mbox{
      \subfigure[Error in $\overline{P_t}$ vs. $t$ for ANISE.]{\epsfig{file=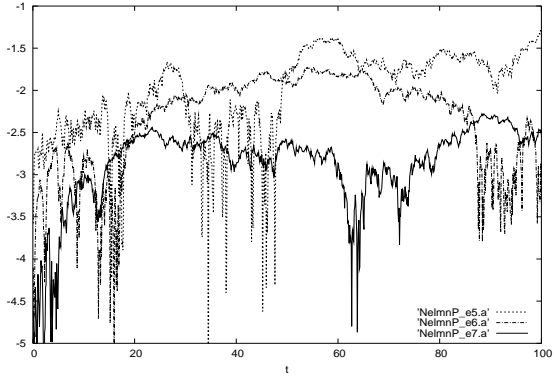,width=3in,height=2in}} \quad
      \subfigure[Error in $\overline{P_t}$ vs. $t$ for SDE9.]{\epsfig{file=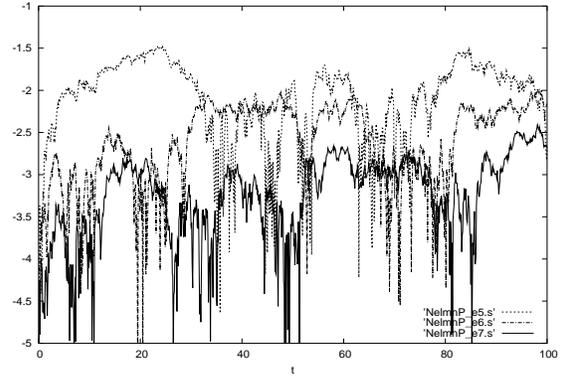,width=3in,height=2in}}
      }
    \mbox{
      \subfigure[Error in ${\rm var}(P_t)$ vs. $t$ for ANISE.]{\epsfig{file=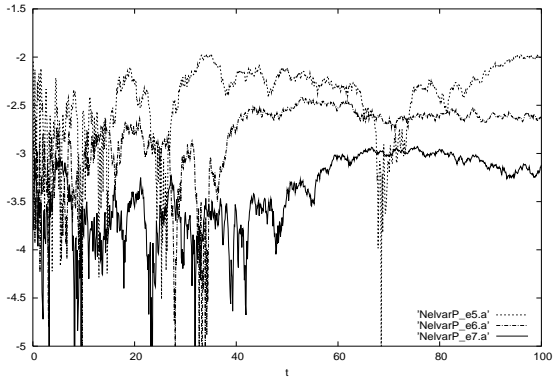,width=3in,height=2in}} \quad
      \subfigure[Error in ${\rm var}(P_t)$ vs. $t$ for SDE9.]{\epsfig{file=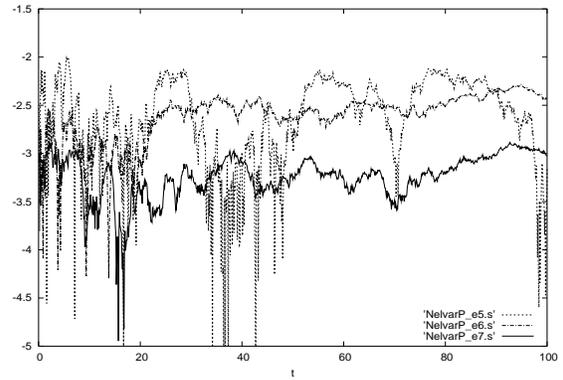,width=3in,height=2in}} 
      }
    \caption{Error in mean and variance of $ P_t$ for Nelson model.}
    \label{Nel1}
  \end{center}
\end{figure}

In Fig. \ref{Nel1} we plot the log base ten mean relative error in mean price $\overline{ P_t}$ 
against time for ANISE (in part (a)) and SDE9 (in part (b)). Results are shown for $10^5$ (dashed curve), $10^6$ (dot-dashed curve) and 
$10^7$ (solid curve) trajectories. Convergence with increasing numbers of trajectories is good in both cases. For ten million 
trajectories the averages have a relative accuracy of about one part in a thousand. Errors this small 
are not visible in plots and so we do not show the actual solutions. The requirement of millions to tens of millions of
trajectories for full convergence is typical of systems of SODEs with white noises, since Monte Carlo error bounds scale as the 
inverse square root of the number of trajectories.

Figure \ref{Nel1} also shows plots of the log base ten mean relative error in variance of price ${\rm var}(P_t)$
against time for ANISE (in part (c)) and SDE9 (in part (d)). Here the convergence is slightly better than that
for mean price.

\begin{figure}[h]
  \begin{center}
    \mbox{
      \subfigure[Error in $\overline{V_t}$ vs. $t$ for ANISE.]{\epsfig{file=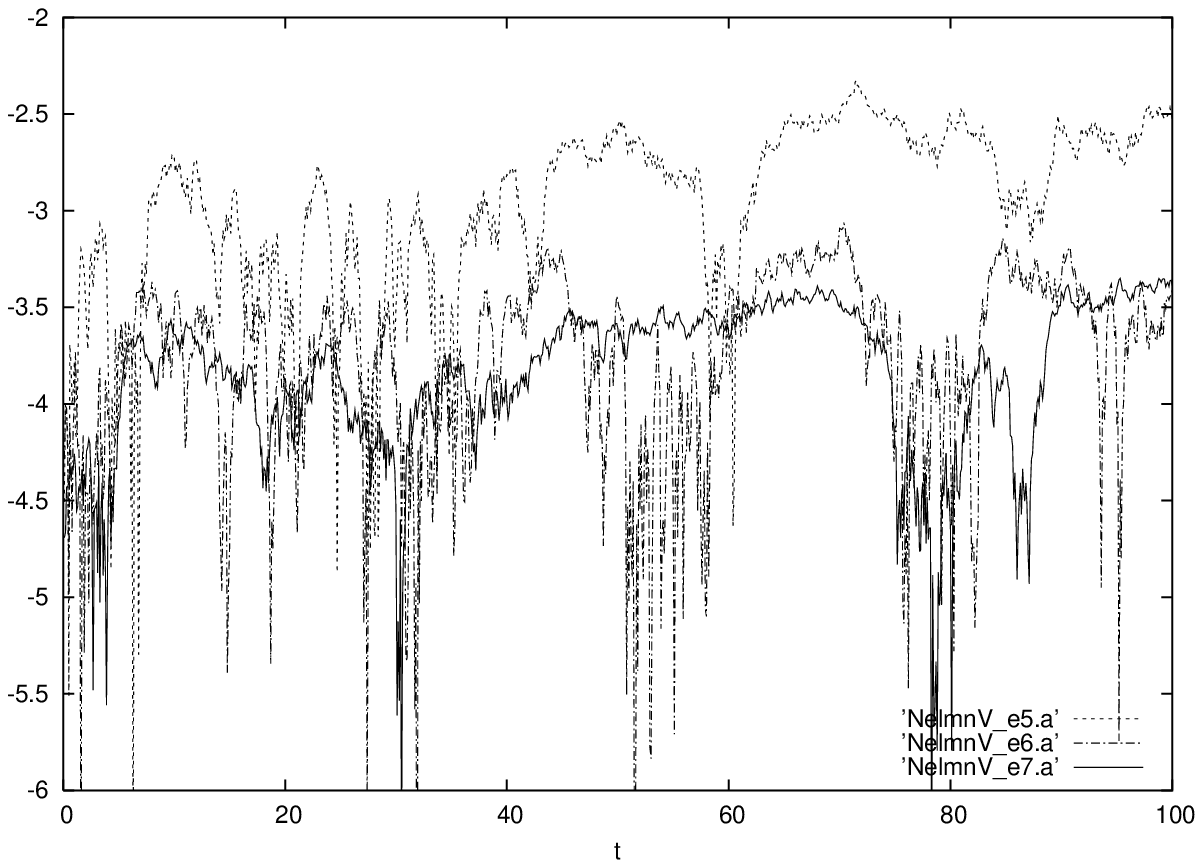,width=3in,height=2in}} \quad
      \subfigure[Error in $\overline{V_t}$ vs. $t$ for SDE9.]{\epsfig{file=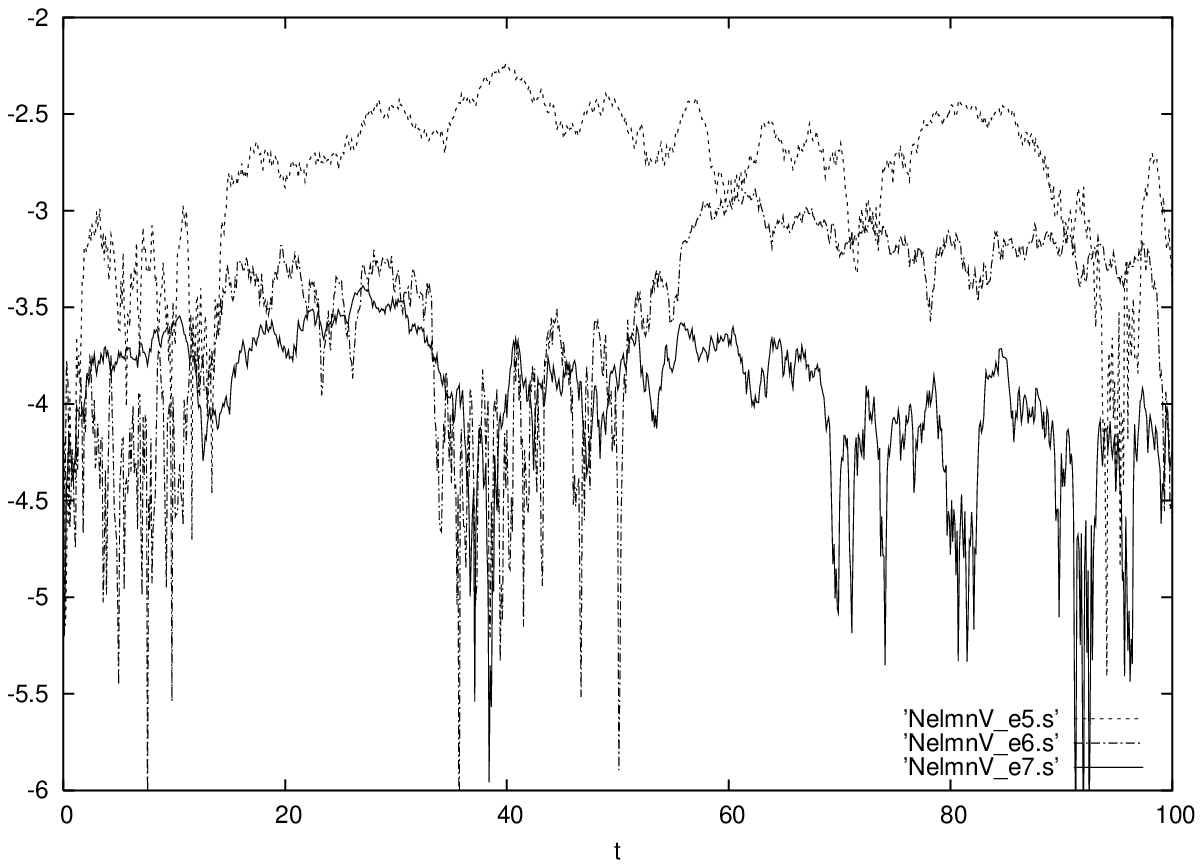,width=3in,height=2in}}
      }
    \mbox{
      \subfigure[Error in ${\rm var}(V_t)$ vs. $t$ for ANISE.]{\epsfig{file=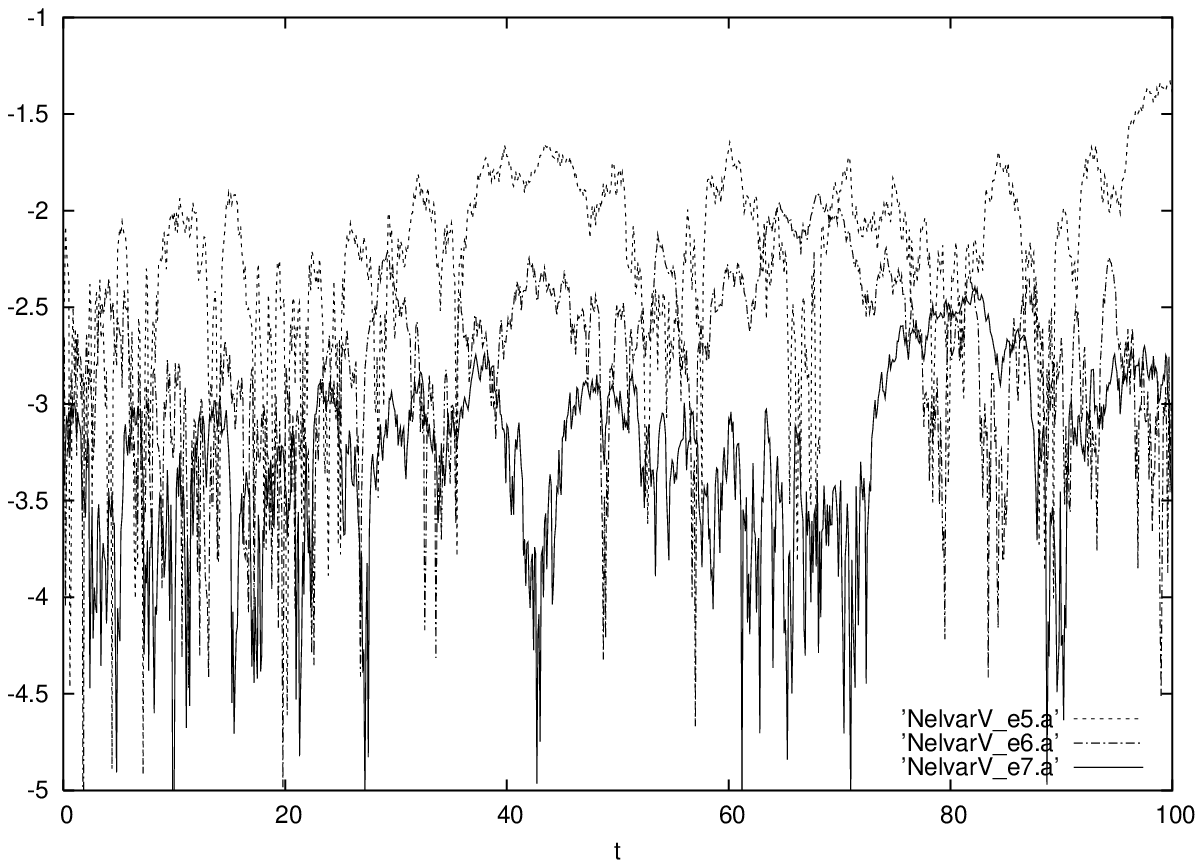,width=3in,height=2in}} \quad
      \subfigure[Error in ${\rm var}(V_t)$ vs. $t$ for SDE9.]{\epsfig{file=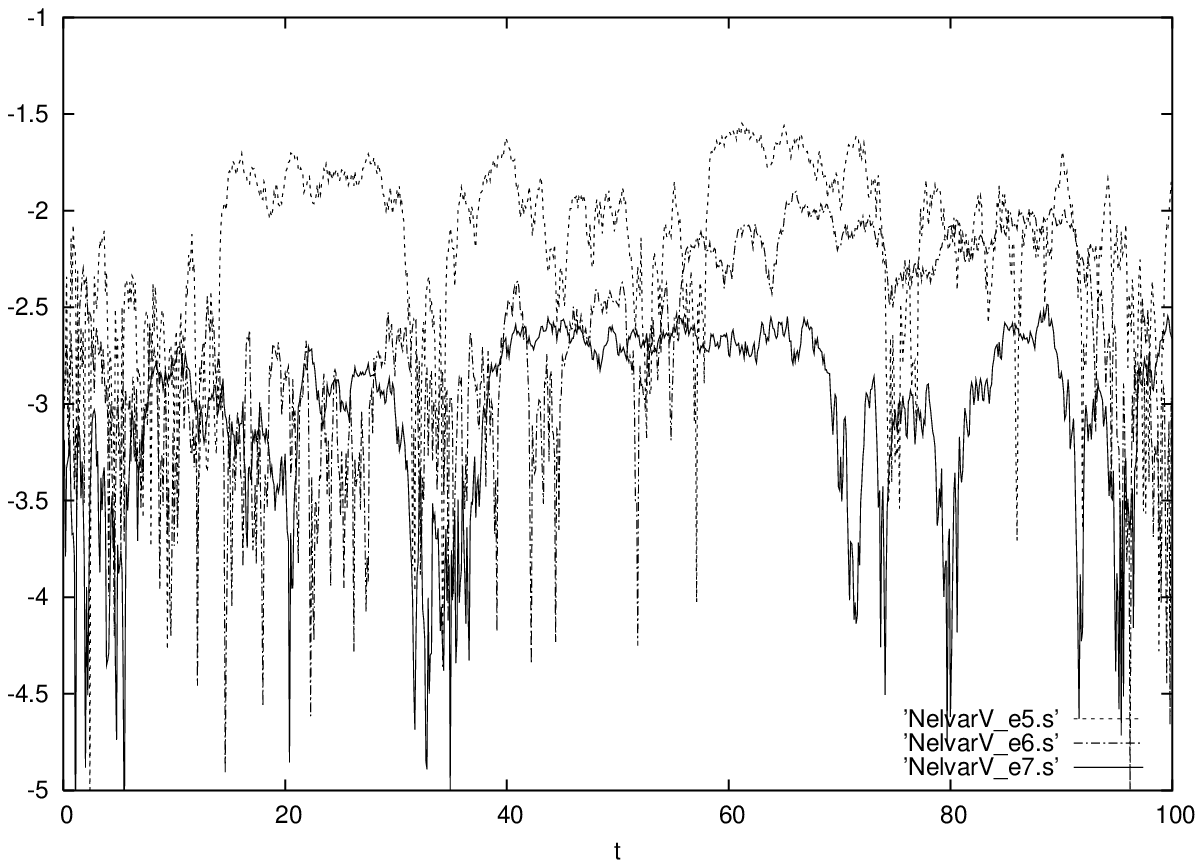,width=3in,height=2in}} 
      }
    \caption{Error in mean and variance of $V_t$ for Nelson model.}
    \label{Nel2}
  \end{center}
\end{figure}

In Fig. \ref{Nel2} we show plots of the log base ten mean relative errors in the mean and variance of the
volatility $V_t$. Here the variance has larger error than the mean. Again ANISE and SDE9 show similar rates of convergence.

The CPU times for various numbers of trajectories are shown in Table \ref{table:T1}. ANISE takes about 7 seconds to compute 1000 trajectories. The SDE9 calculations take about 60\% longer.
\begin{table}[h]
\begin{center}
\begin{tabular}{|c|c|c|c|}
\hline
~ \# Trajectories ~ &~~ ANISE CPU Time ~~&~~ SDE9 CPU Time ~~&~~ CPU Time Ratio SDE9/ANISE ~~ \\ \hline
~ $10^3$ & 0.69E+01 & 0.11E+02 & 1.59 \\ ~ $10^4$ & 0.68E+02 & 0.11E+03 & 1.62 \\~ $10^5$ & 0.68E+03 & 0.11E+04 & 1.63 \\~ $10^6$ & 0.68E+04 & 0.11E+05 & 1.62 \\ ~ $10^7$ & 0.68E+05 & 0.11E+06 & 1.62 \\
\hline
\end{tabular}
\end{center}
\caption{CPU times for Nelson Model in seconds.}
\label{table:T1}
\end{table}
This table also shows that both methods scale well with the number of trajectories. In other words there are no rare
problematic trajectories. 

\subsection{Hull-White Model}

For the Hull-White model\cite{Davis,HW,Hull} the log-price $P_t$ and volatility $V_t$ obey SODEs
\begin{eqnarray}
dP_t&=& \sqrt{V_t} dW_{1t}\nonumber \\
dV_t&=&(\theta-\lambda V_t) dt+\gamma dW_{2t}.\nonumber
\end{eqnarray}
Thus, we again have two equations and two Wiener processes.

The derivatives needed by the numerical methods are given in Table \ref{table:D2}. A time step of $dt=.1$ was used and the
equations were integrated to 100.
\begin{table}[h]
\begin{center}
\begin{tabular}{c|ccc}
~ $X_t$ ~ & ~ $\frac{\partial X_t}{\partial t}$ ~ & ~ $\frac{\partial X_t}{\partial W_{1t}}$ ~ & ~ $\frac{\partial X_t}{\partial W_{2t}}$ ~ \\ \hline 
 $P_t$ & 0 & $\sqrt{V_t}$ & 0 \\ $V_t$ & ~~$\theta-\lambda V_t$ & 0 & $\gamma$ \\ 
\end{tabular}
\end{center}
\caption{ Derivatives for Hull-White Model. }
\label{table:D2}
\end{table}
Once again care must be taken to program $\sqrt{|V_t|}$ rather than $\sqrt{V_t}$.

We explored convergence for the average quantities $\overline{ P_t}$, ${\rm var}(P_t)$, $\overline{V_t}$, and ${\rm var}(V_t)$. The exact solutions for these observables are given by
\begin{eqnarray}
\overline{ P_t}&=&P_0\\
{\rm var}(P_t)&=&\frac{V_0}{\lambda}(1-e^{-\lambda t})+\frac{\theta}{\lambda^2}(\lambda t+e^{-\lambda t}-1)\\
 \overline{V_t}&=&V_0e^{-\lambda t}+\frac{\theta}{\lambda}(1-e^{-\lambda t})\\
{\rm var}(V_t)&=&\frac{\gamma^2}{2\lambda}(1-e^{-2\lambda t})
\end{eqnarray}
for this model.

We set parameters $\theta=.03$, $\lambda=.035$, $\gamma=.0068$, and initial conditions $V_0=.029$ and $P_0=.5$.

\begin{figure}[h]
  \begin{center}
    \mbox{
      \subfigure[Error in $\overline{P_t}$ vs. $t$ for ANISE.]{\epsfig{file=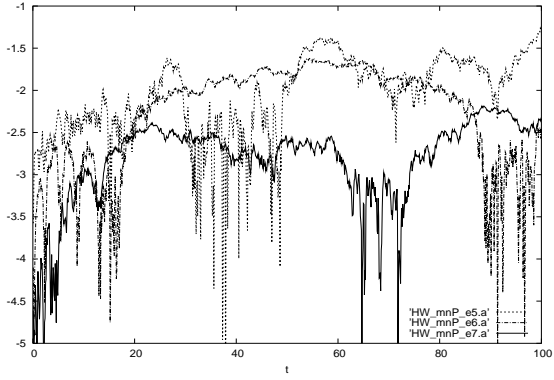,width=3in,height=2in}} \quad
      \subfigure[Error in $\overline{P_t}$ vs. $t$ for SDE9.]{\epsfig{file=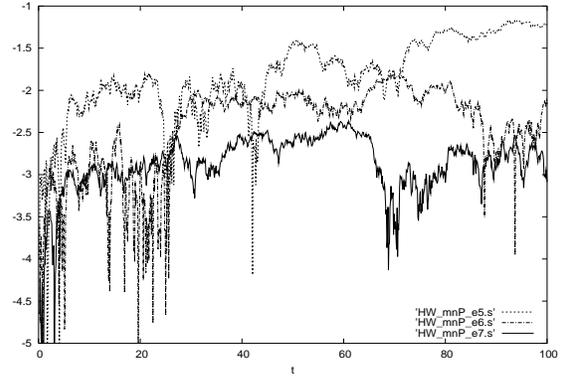,width=3in,height=2in}}
      }
    \mbox{
      \subfigure[Error in ${\rm var}(P_t)$ vs. $t$ for ANISE.]{\epsfig{file=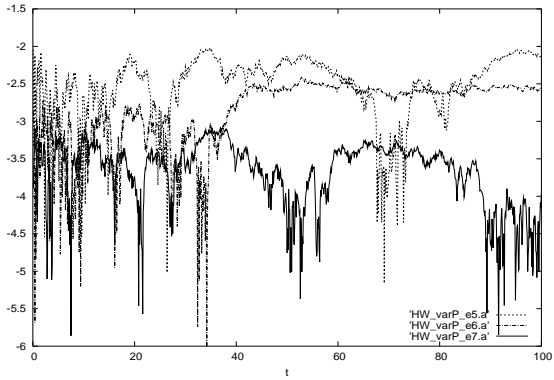,width=3in,height=2in}} \quad
      \subfigure[Error in ${\rm var}(P_t)$ vs. $t$ for SDE9.]{\epsfig{file=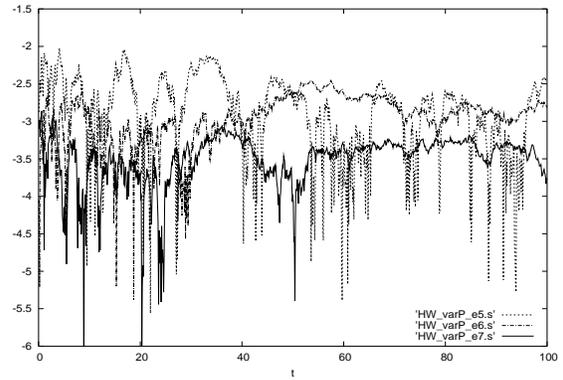,width=3in,height=2in}} 
      }
    \caption{Error in mean and variance of $P_t$ for Hull-White model.}
    \label{HW1}
  \end{center}
\end{figure}

In Fig. \ref{HW1} we plot the log base ten error in the mean and variance of $P_t$ for ANISE ((a) and (c), respectively) and for SDE9 ((b) and (d), respectively). Dotted curves show the results for $10^5$ trajectories while dot-dashed and solid curves are for $10^6$ and $10^7$ trajectories, respectively. Good
convergence with numbers of trajectories is seen. Once again, convergence for the variance is slightly better than that for the mean. 

\begin{figure}[h]
  \begin{center}
    \mbox{
      \subfigure[Error in $\overline{V_t}$ vs. $t$ for ANISE.]{\epsfig{file=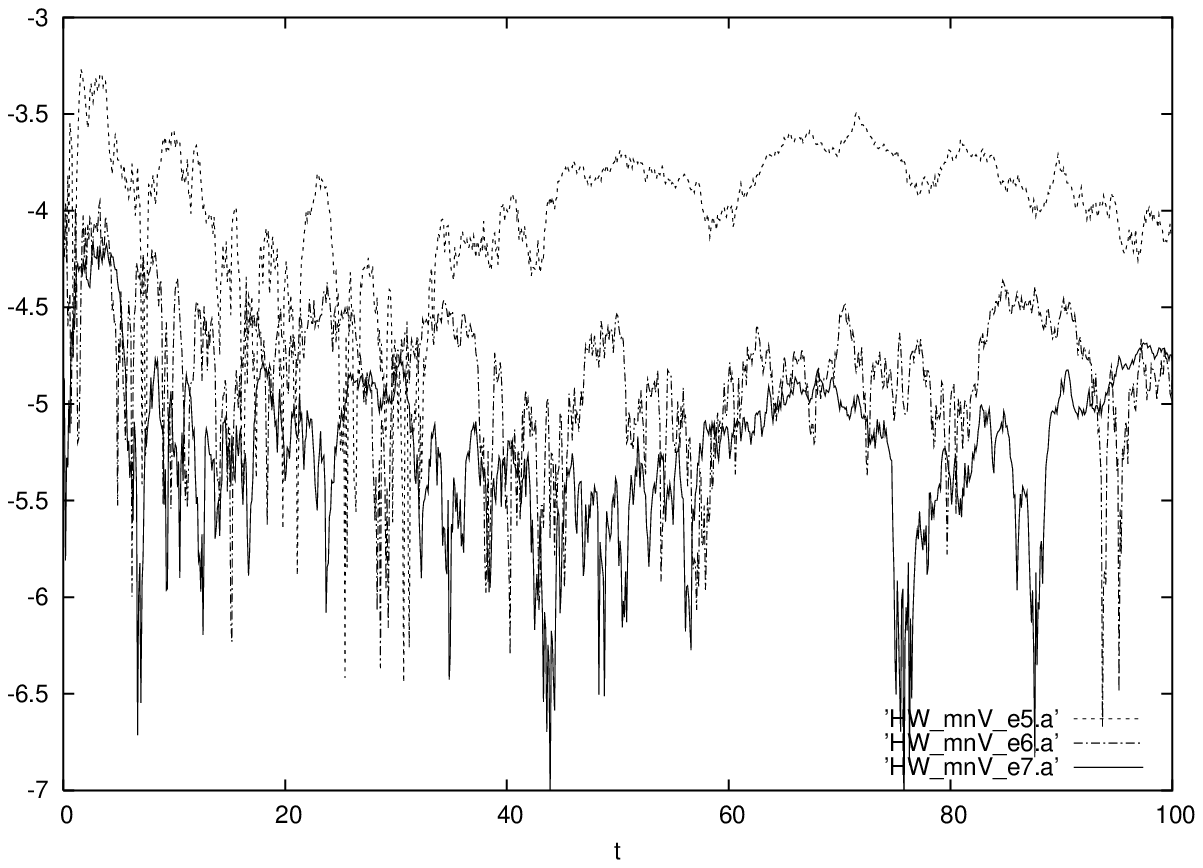,width=3in,height=2in}} \quad
      \subfigure[Error in $\overline{V_t}$ vs. $t$ for SDE9.]{\epsfig{file=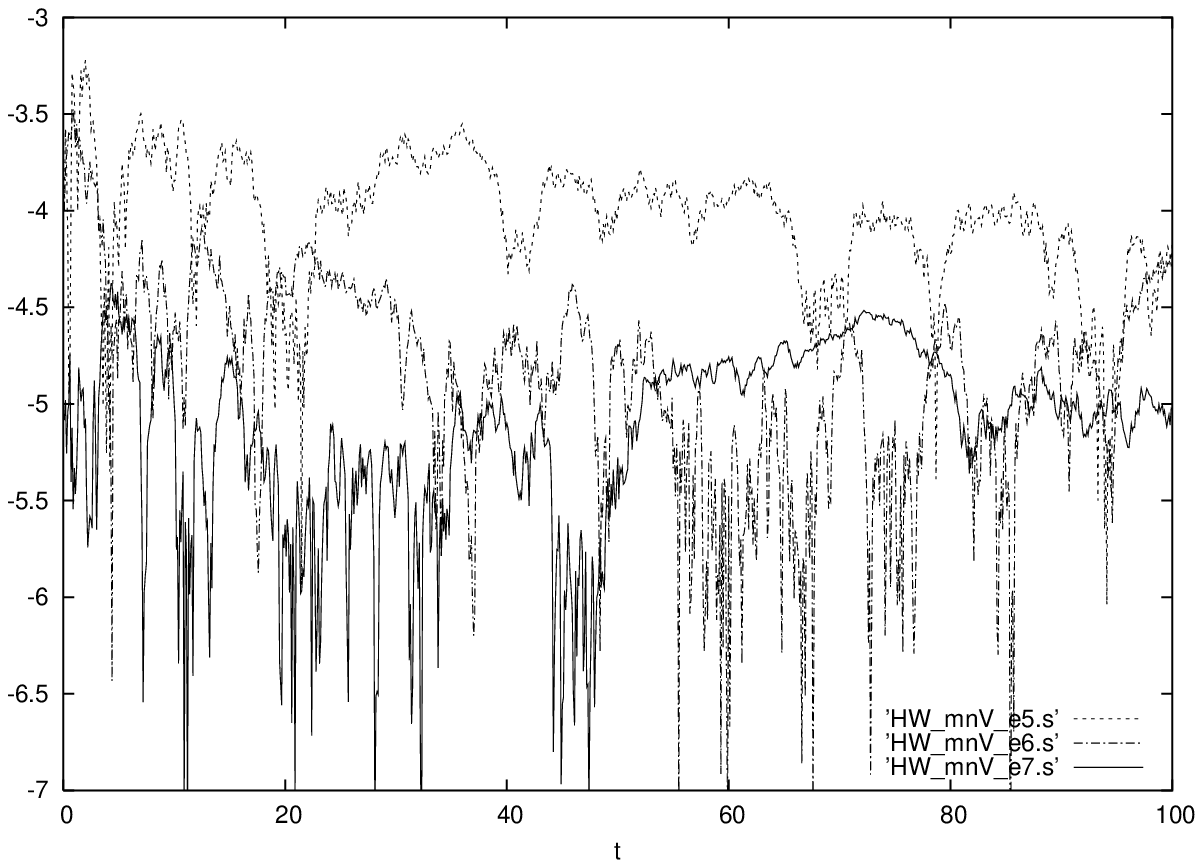,width=3in,height=2in}}
      }
    \mbox{
      \subfigure[Error in ${\rm var}(V_t)$ vs. $t$ for ANISE.]{\epsfig{file=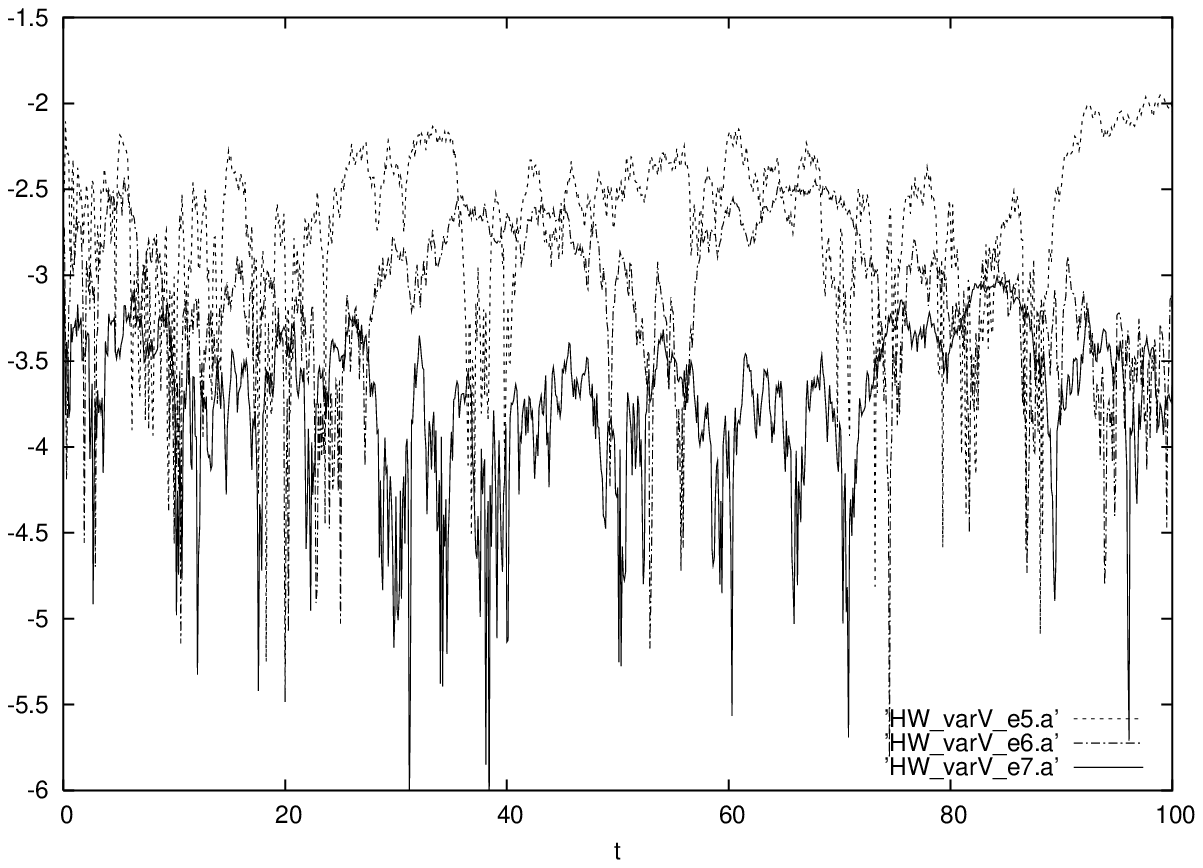,width=3in,height=2in}} \quad
      \subfigure[Error in ${\rm var}(V_t)$ vs. $t$ for SDE9.]{\epsfig{file=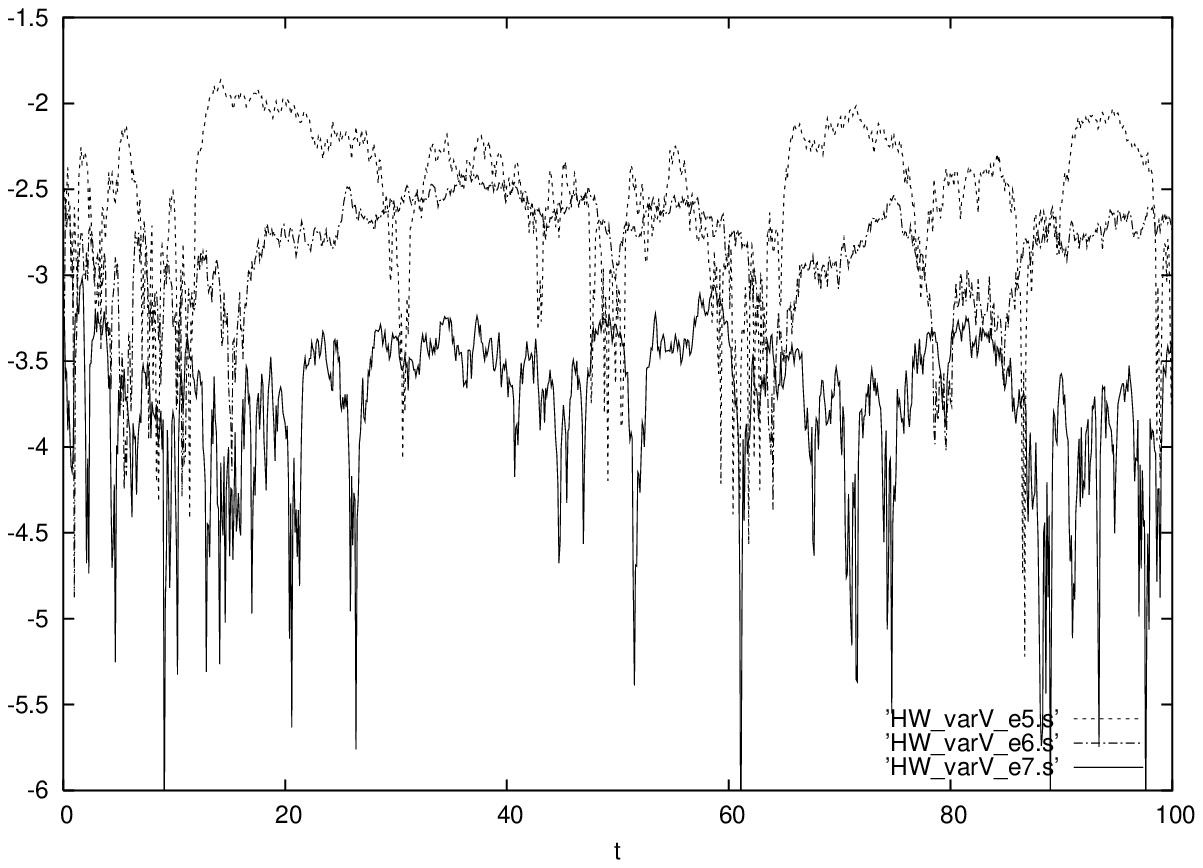,width=3in,height=2in}} 
      }
    \caption{Error in mean and variance of $V_t$ for Hull-White model.}
    \label{HW2}
  \end{center}
\end{figure}

Figure \ref{HW2} shows the log base ten error in the mean and variance of the volatility $V_t$ for ANISE ((a) and (c), respectively) and for SDE9 ((b) and (d), respectively). Again good convergence to the exact results is observed. The errors in the variance are larger than those in the mean.

Cpu times are compared in Table \ref{table:T2} for various numbers of trajectories. ANISE takes 7.5 s to compute 1000 trajectories. Again we observe that SDE9 takes 50 \% longer.
\begin{table}[h]
\begin{center}
\begin{tabular}{|c|c|c|c|}
\hline
~ \# Trajectories ~ &~~ ANISE CPU Time ~~&~~ SDE9 CPU Time ~~&~~ CPU Time Ratio SDE9/ANISE ~~ \\ \hline
~ $10^3$  & 0.76E+01 & 0.11E+02 & 1.47 \\ ~ $10^4$ & 0.76E+02 & 0.11E+03 & 1.44 \\~ $10^5$ & 0.76E+03 & 0.11E+04 & 1.46 \\~ $10^6$ & 0.76E+04 & 0.11E+05 & 1.46 \\ ~ $10^7$ & 0.76E+05 & 0.11E+06 & 1.46 \\
\hline
\end{tabular}
\end{center}
\caption{CPU times for Hull-White in seconds.}
\label{table:T2}
\end{table}
Once again good scaling is obtained for both methods with the number of trajectories.

\subsection{Cox-Ingersoll-Ross Model}

The SODEs for the Cox-Ingersoll-Ross model\cite{CIR} are
\begin{eqnarray}
dP_t&=& (\alpha dt+\sqrt{V_t}dW_{1t})P_t\\
dV_t&=&\kappa (\theta-V_t) dt+\sigma \sqrt{V_t} ~[\rho dW_{1t}+\sqrt{1-\rho^2}dW_{2t}]
\end{eqnarray}
and so we have two equations with two Wiener processes. In this case the volatility depends on both Wiener processes.

The derivatives needed by the numerical methods are provided in Table \ref{table:D3}. A time step of $dt=.01$ was used and the equations were
integrated to 10.
\begin{table}[h]
\begin{center}
%\begin{tabular}{|c|c|c|c|c|c|}
\begin{tabular}{c|ccc}
~ $X_t$ ~ & ~ $\frac{\partial X_t}{\partial t}$ ~ & ~ $\frac{\partial X_t}{\partial W_{1t}}$ ~ & ~ $\frac{\partial X_t}{\partial W_{2t}}$ ~ \\ \hline 
 $P_t$ & ~~$[\alpha -\frac{1}{2}(V_t+\frac{1}{2}\sigma\rho)] P_t$~~ & $P_t\sqrt{V_t}$ & 0 \\ $V_t$ & $\kappa (\theta-V_t)-\frac{1}{4}\sigma^2$ & $\sigma\rho \sqrt{V_t}$ & ~~$\sigma\sqrt{1-\rho^2}\sqrt{V_t}$ \\ 
\end{tabular}
\end{center}
\caption{ Derivatives for Cox-Ingersoll-Ross Model. }
\label{table:D3}
\end{table}

We look for convergence in four observables; mean log-price $\overline{\ln P_t}$, variance in log-price ${\rm var}(\ln P_t)$, mean volatility $\overline{V_t}$, and variance of the volatility ${\rm var}(V_t)$. The exact solutions for these quantities are given by
\begin{eqnarray}
\overline{\ln P_t}&=&\ln P_0+(\alpha-\frac{\theta}{2})t+\frac{1}{2\kappa}(V_0-\theta)(e^{-\kappa t}-1)\\
{\rm var}(\ln P_t)&=&[\theta-\frac{\sigma\theta}{\kappa}(\rho-\frac{\sigma}{4\kappa})]t+\frac{\sigma}{\kappa}(V_0-\theta)(\rho-\frac{\sigma}{2\kappa})te^{-\kappa t}\nonumber \\
&+&\{\frac{\sigma}{\kappa^2}[(V_0-\theta)(\rho-\frac{\sigma}{2\kappa})-\theta(\rho-\frac{\sigma}{4\kappa})]-\frac{V_0-\theta}{\kappa}\}(e^{-\kappa t}-1)\nonumber \\
&-&\frac{\sigma^2}{4\kappa^3}(V_0-\frac{\theta}{2})(e^{-\kappa t}-1)^2\\
\overline{V_t}&=&V_0e^{-\kappa t}+\theta (1-e^{-\kappa t})\\
{\rm var}(V_t)&=&\frac{V_0\sigma^2}{\kappa}(e^{-\kappa t}-e^{-2\kappa t})+ \frac{\theta\sigma^2}{2\kappa} (1-e^{-\kappa t})^2
\end{eqnarray}
The parameters were set to $\alpha=.1$, $\kappa=.29368$, $\theta=.07935$, $\sigma=.11425$, $\rho=-.2$ and price and 
volatility was set to initial values $P_0=1$ and $V_0=.1$. 

\begin{figure}[htbp]
  \begin{center}
    \mbox{
      \subfigure[Error in $\overline{\ln P_t}$ vs. $t$ for ANISE.]{\epsfig{file=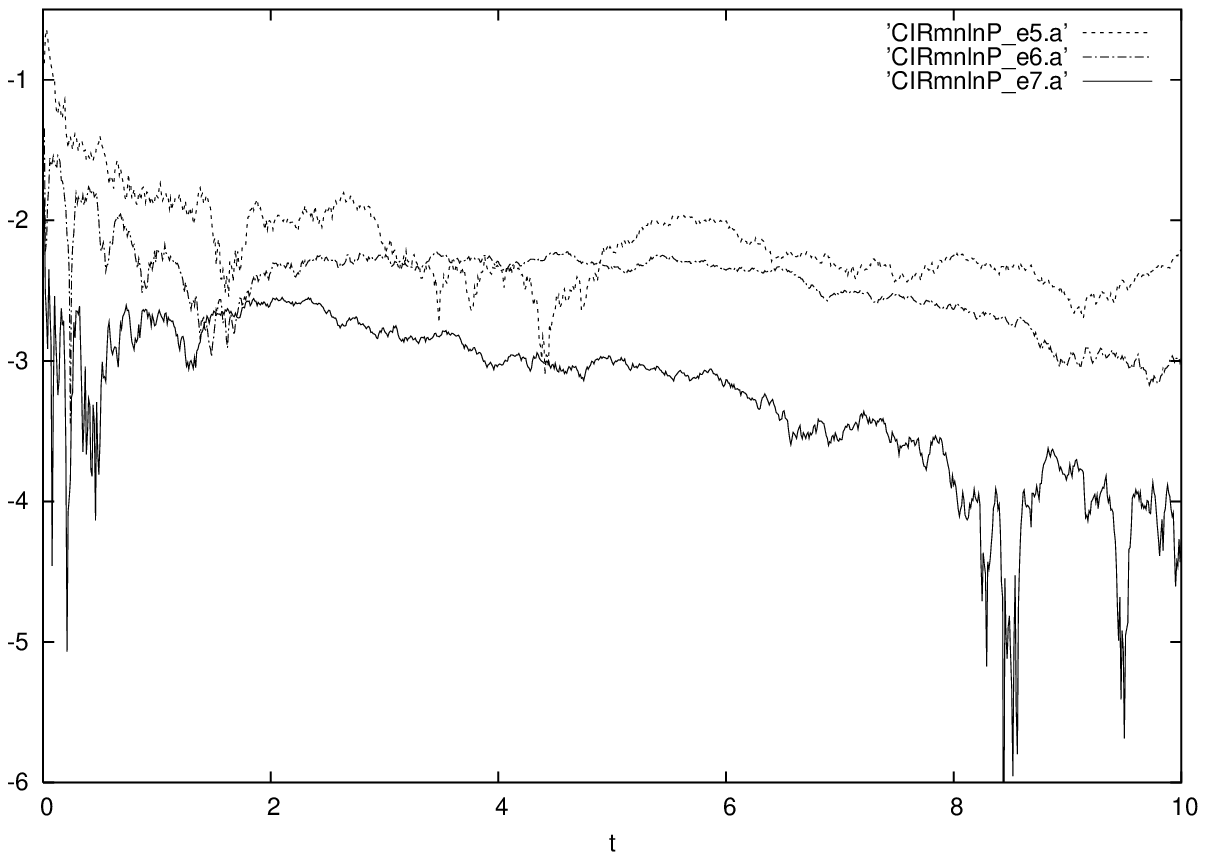,width=3in,height=2in}} \quad
      \subfigure[Error in $\overline{\ln P_t}$ vs. $t$ for SDE9.]{\epsfig{file=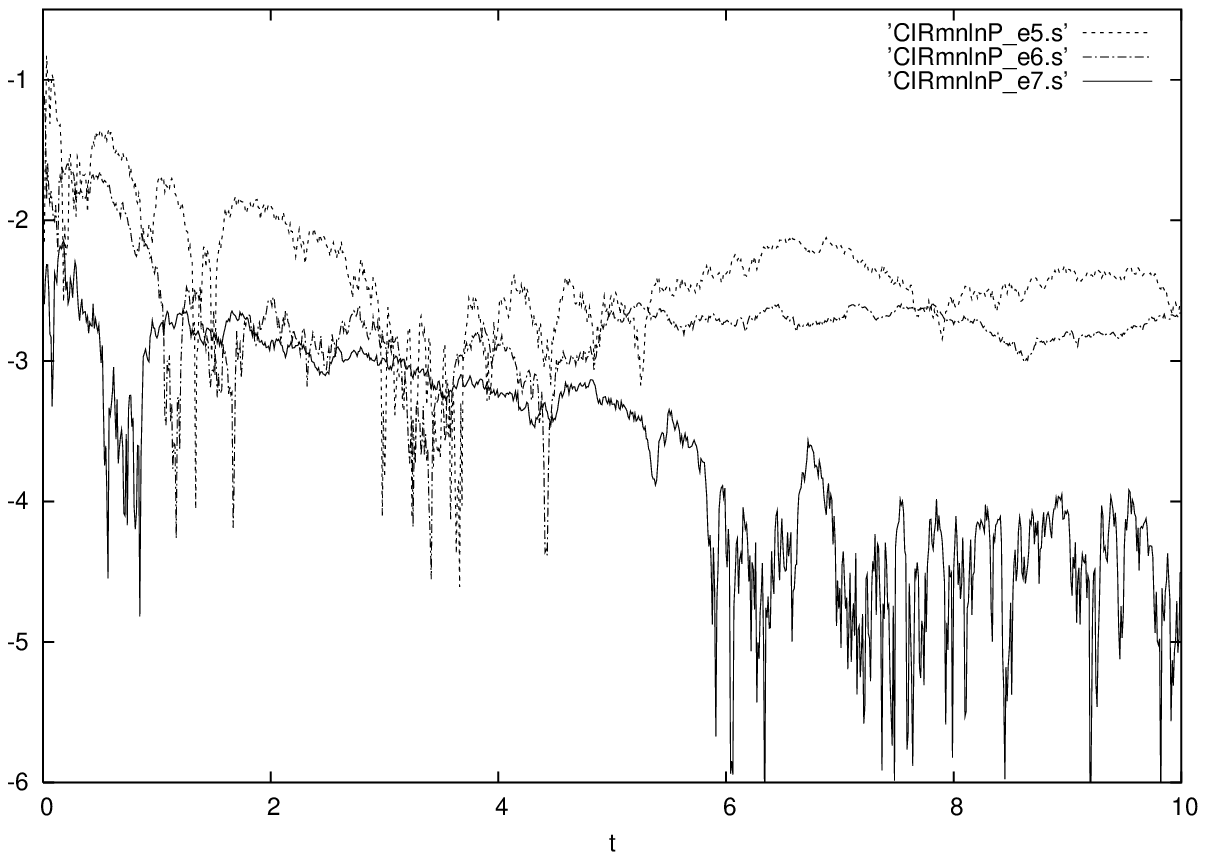,width=3in,height=2in}}
      }
    \mbox{
      \subfigure[Error in ${\rm var}(\ln P_t)$ vs. $t$ for ANISE.]{\epsfig{file=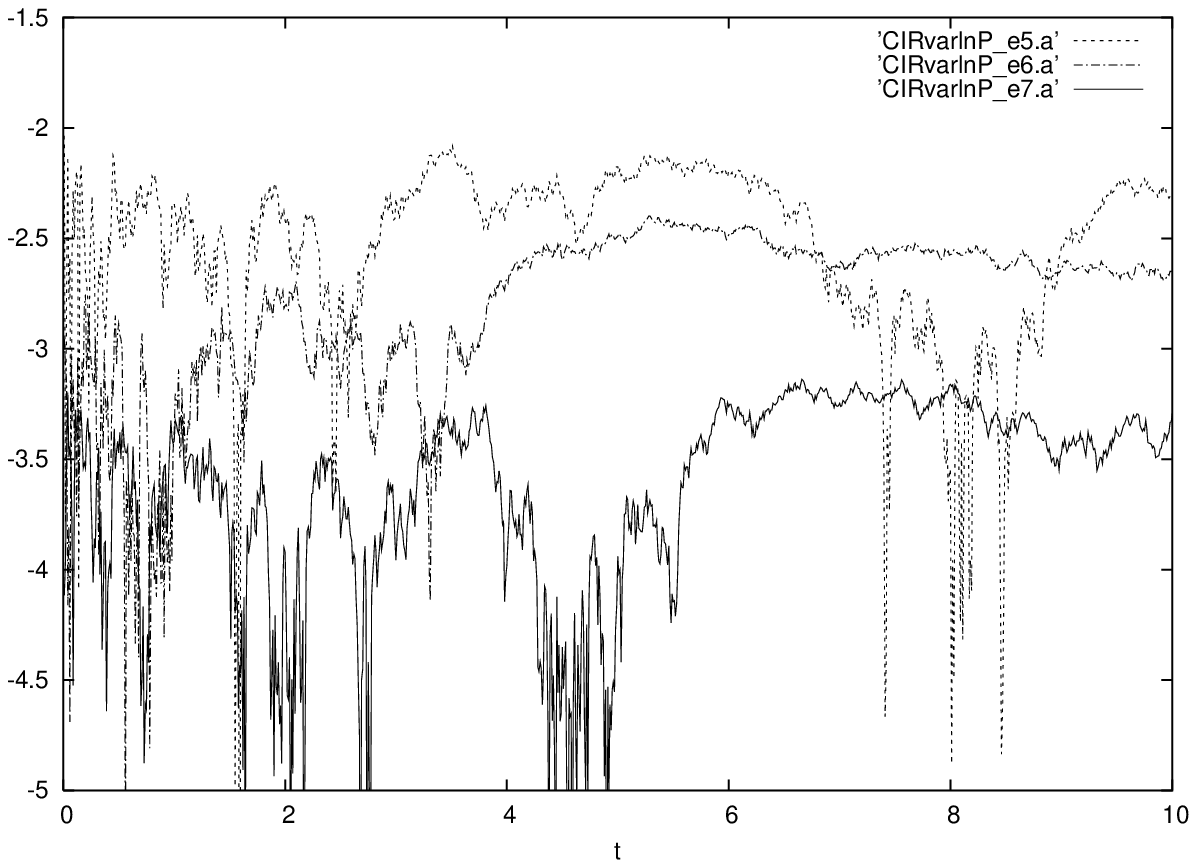,width=3in,height=2in}} \quad
      \subfigure[Error in ${\rm var}(\ln P_t)$ vs. $t$ for SDE9.]{\epsfig{file=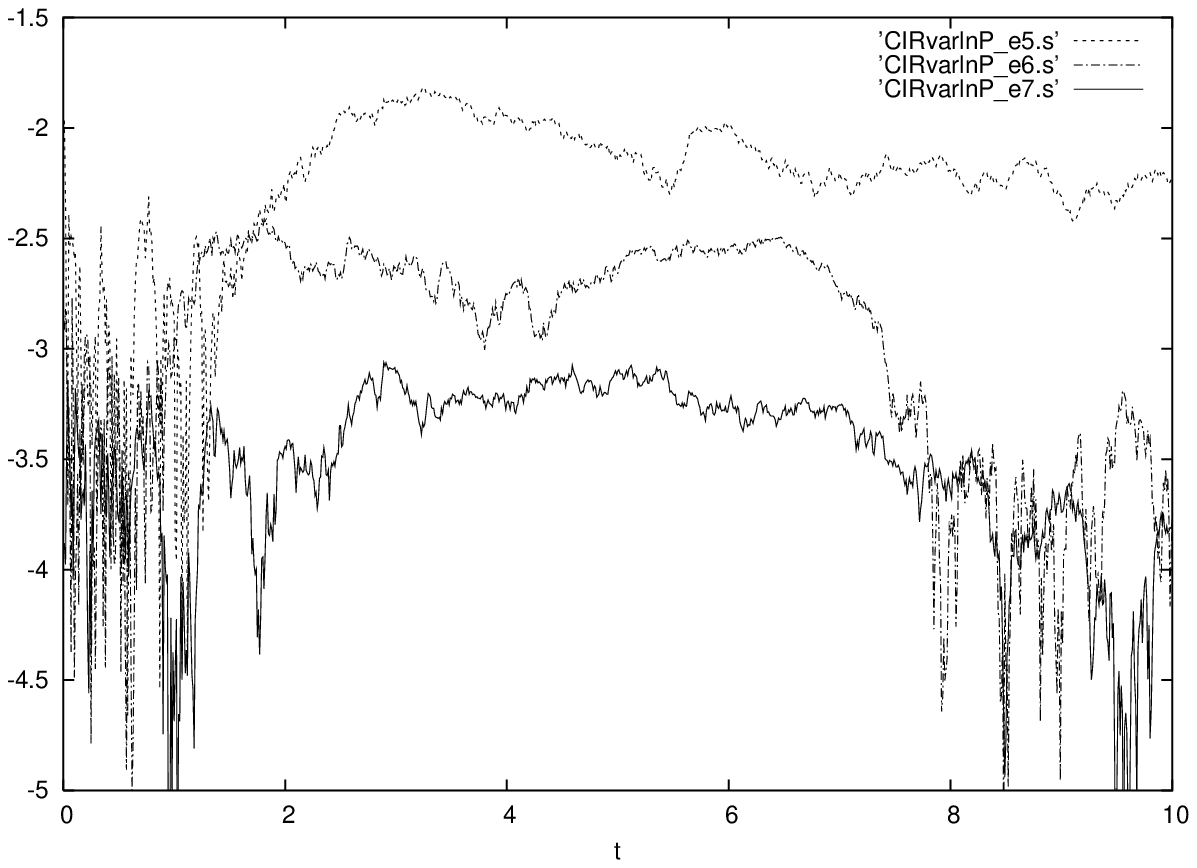,width=3in,height=2in}} 
      }
    \caption{Error in mean and variance of $\ln P_t$ for Cox-Ingersoll-Ross model.}
    \label{CIR1}
  \end{center}
\end{figure}

In Fig. \ref{CIR1} we show the log base ten error in $\overline{\ln P_t}$ and ${\rm var}(\ln P_t)$ plotted against time 
for ANISE ((a) and (c)) and SDE9 ((b) and (d)). Dashed, dot-dashed and solid curves represent errors for runs of $10^5$, $10^6$ and $10^7$
trajectories, respectively. Good convergence is seen for both methods at all times, although errors in the variance are larger
than those in the mean.

\begin{figure}[htbp]
  \begin{center}
    \mbox{
      \subfigure[Error in $\overline{V_t}$ vs. $t$ for ANISE.]{\epsfig{file=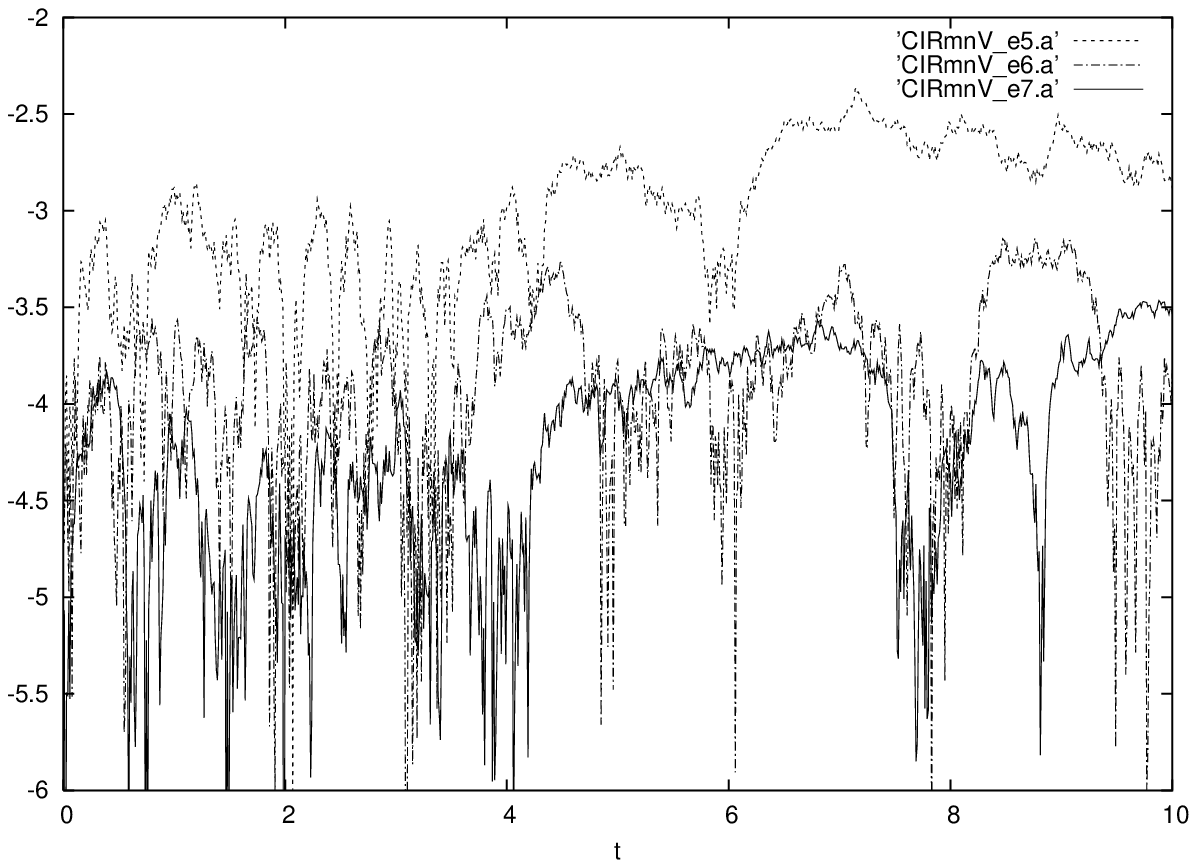,width=3in,height=2in}} \quad
      \subfigure[Error in $\overline{V_t}$ vs. $t$ for SDE9.]{\epsfig{file=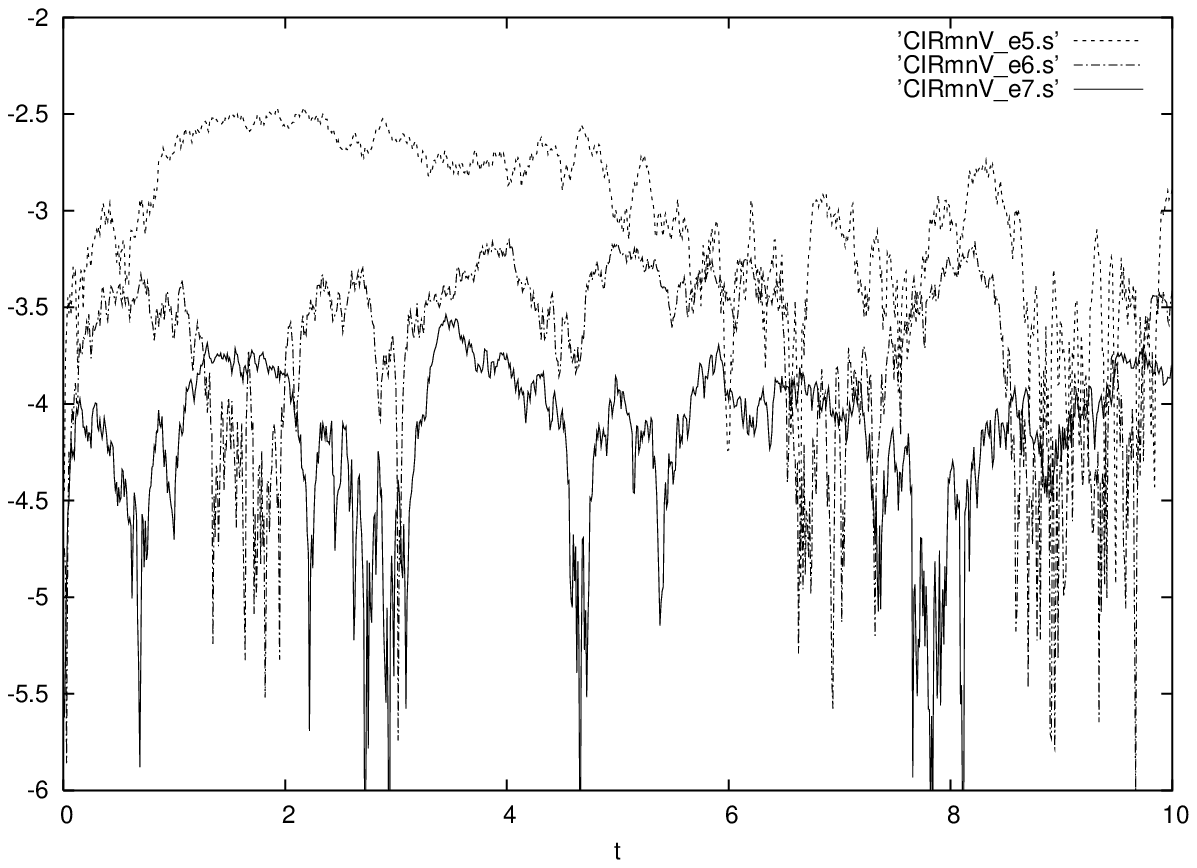,width=3in,height=2in}}
      }
    \mbox{
      \subfigure[Error in ${\rm var}(V_t)$ vs. $t$ for ANISE.]{\epsfig{file=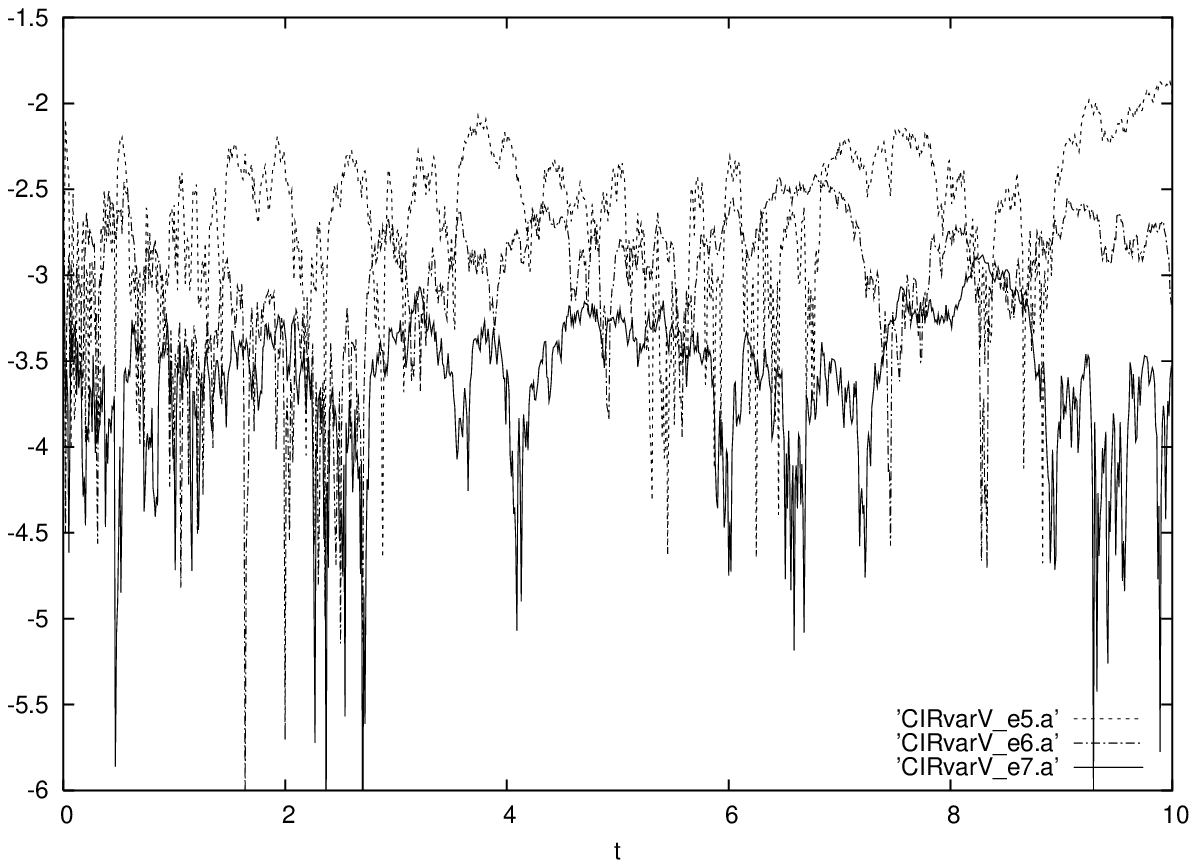,width=3in,height=2in}} \quad
      \subfigure[Error in ${\rm var}(V_t)$ vs. $t$ for SDE9.]{\epsfig{file=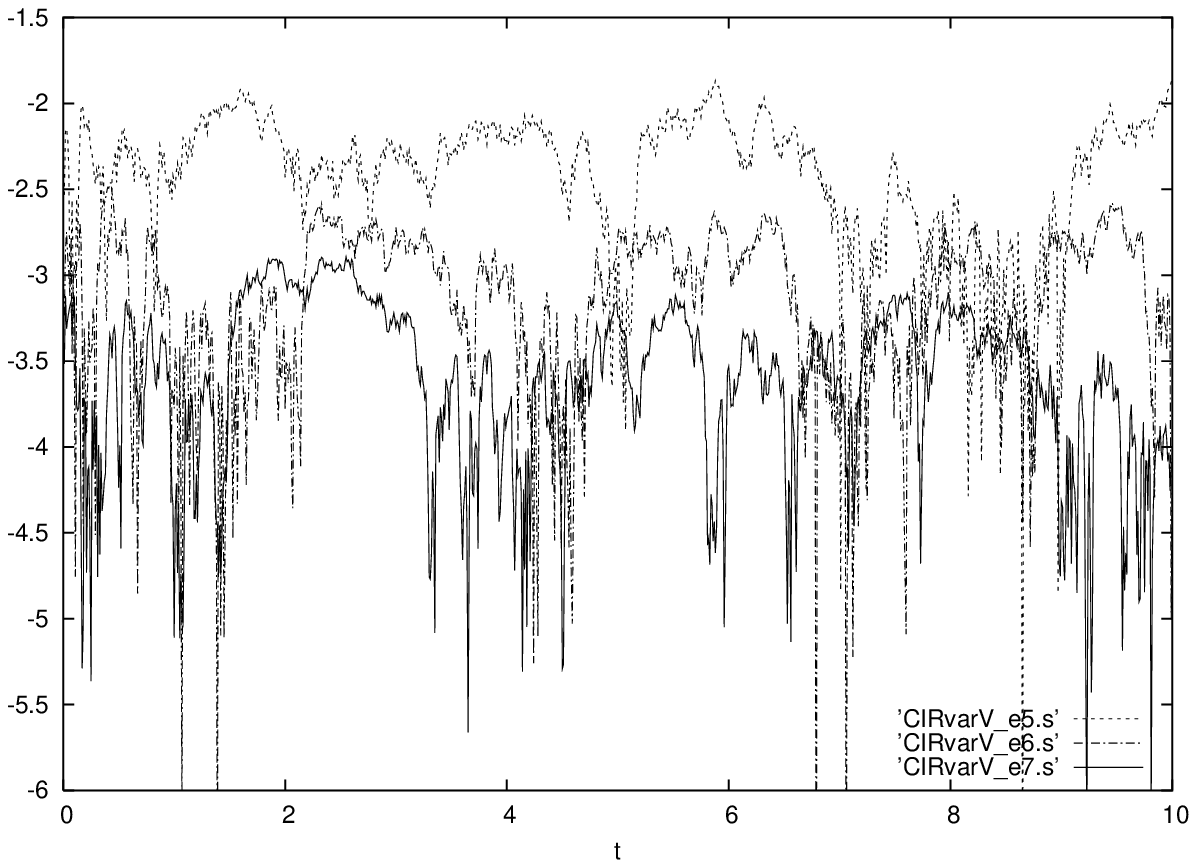,width=3in,height=2in}} 
      }
    \caption{Error in mean and variance of $V_t$ for Cox-Ingersoll-Ross model.}
    \label{CIR2}
  \end{center}
\end{figure}

Figure \ref{CIR2} plots errors in $\overline{V_t}$ and ${\rm var}(V_t)$. Again we see excellent convergence in both cases. Errors in the variance
are bigger than those in the mean.

The cpu times for various numbers of trajectories are shown in Table \ref{table:T3}. ANISE takes less than 5 s to compute 1000 trajectories. The ratio of cpu time for SDE9 to that of ANISE
is now a much larger 2.5. This relative  slowing down of SDE9 is probably caused by the fact that $V_t$ now depends on two Wiener
processes. Once again the ratio is independent of the number of trajectories indicating that both methods handle all trajectories
equally well.
\begin{table}[h]
\begin{center}
\begin{tabular}{|c|c|c|c|}
\hline
~ \# Trajectories ~ &~~ ANISE CPU Time ~~&~~ SDE9 CPU Time ~~&~~ CPU Time Ratio SDE9/ANISE ~~ \\ \hline
~ $10^3$  & 0.48E+01 & 0.12E+02 & 2.45 \\ ~ $10^4$ & 0.48E+02 & 0.12E+03 & 2.46 \\~ $10^5$ & 0.48E+03 & 0.12E+04 & 2.46 \\~ $10^6$ & 0.48E+04 & 0.12E+05 & 2.47 \\ ~ $10^7$ & 0.48E+05 & 0.12E+06 & 2.47 \\
\hline
\end{tabular}
\end{center}
\caption{CPU times for Cox-Ingersoll-Ross model in seconds.}
\label{table:T3}
\end{table}

\subsection{Log-Ornstein-Uhlenbeck Model}

The fourth example is the Log Ornstein-Uhlenbeck model\cite{Scott} for price $P_t$ and volatility $V_t$. In this model
\begin{eqnarray}
dP_t&=&(a dt+e^{V_t}dW_{1t} )P_t\nonumber \\
dV_t&=&(a-b V_t) dt+\frac{1}{2}[\rho dW_{1t} +\sqrt{1-\rho^2}dW_{2t}]\nonumber
\end{eqnarray}
and so we again have two equations and two Wiener processes. The volatility depends on both Wiener processes.

The derivatives required by the SODE methods are given in Table \ref{table:D4}. A time step of $10^{-4}$ was used and the equations were
integrated to 0.1.
\begin{table}[h]
\begin{center}
\begin{tabular}{c|ccc}
~ $X_t$ ~ & ~ $\frac{\partial X_t}{\partial t}$ ~ & ~ $\frac{\partial X_t}{\partial W_{1t}}$ ~ & ~ $\frac{\partial X_t}{\partial W_{2t}}$ ~ \\ \hline 
 $P_t$ & ~~$a P_t-\frac{1}{2}P_t e^{V_t}(\frac{1}{2}\rho+e^{V_t})$~~ & $e^{V_t}P_t$ & 0 \\ $V_t$ & $a+bV_t$ & $\frac{1}{2}\rho$ & ~~$\frac{1}{2}\sqrt{1-\rho^2}$ \\ 
\end{tabular}
\end{center}
\caption{ Derivatives for Log Ornstein-Uhlenbeck Model. }
\label{table:D4}
\end{table}

We look for convergence in three quantities; mean log-price $\overline{\ln P_t}$, mean volatility $\overline{V_t}$, and variance in volatility ${\rm var}(V_t)$. Exact solutions for these observables are given by
\begin{eqnarray}
\overline{\ln P_t}&=&\ln P_0+at-\frac{1}{2}\int_0^tdt'~\exp\{2[V_0e^{-bt'}+\frac{a}{b}(1-e^{-bt'})]+\frac{1}{4b}(1-e^{-2bt'})\}\\
\overline{V_t}&=&V_0 e^{-bt}+\frac{a}{b}(1-e^{-bt})\\
{\rm var}(V_t)&=&\frac{1}{8b}(1-e^{-2bt})
\end{eqnarray}
The solution for $\overline{\ln P_t}$ was obtained using a variable-stepsize Runge-Kutta code for ODEs\cite{Hair}. Parameters were set to $a=70$, $b=100$, $\rho=.2$ and initial conditions $P_0=.5$
and $V_0=.029$ were used. 

\begin{figure}[h]
  \begin{center}
    \mbox{
      \subfigure[Error in $\overline{P_t}$ vs. $t$ for ANISE.]{\epsfig{file=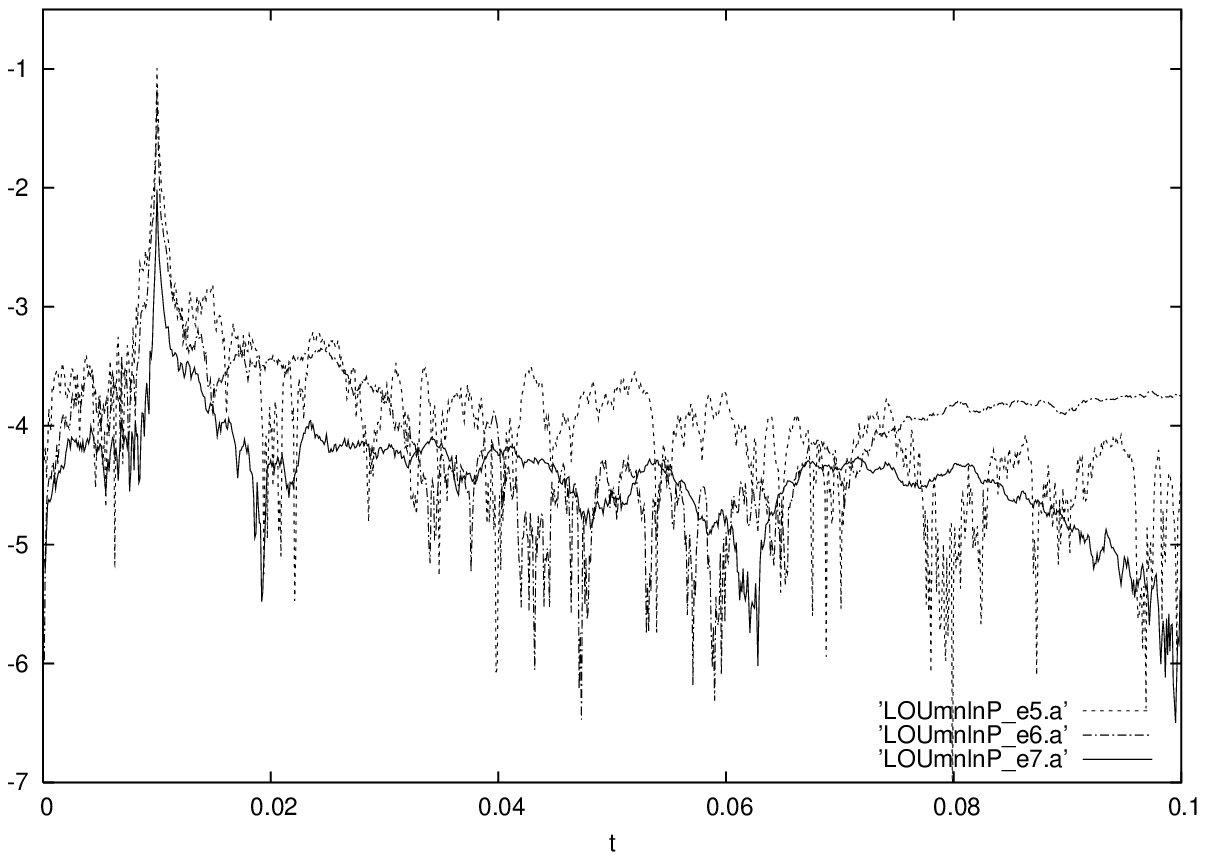,width=3in,height=2in}}
      \subfigure[Error in $\overline{P_t}$ vs. $t$ for SDE9.]{\epsfig{file=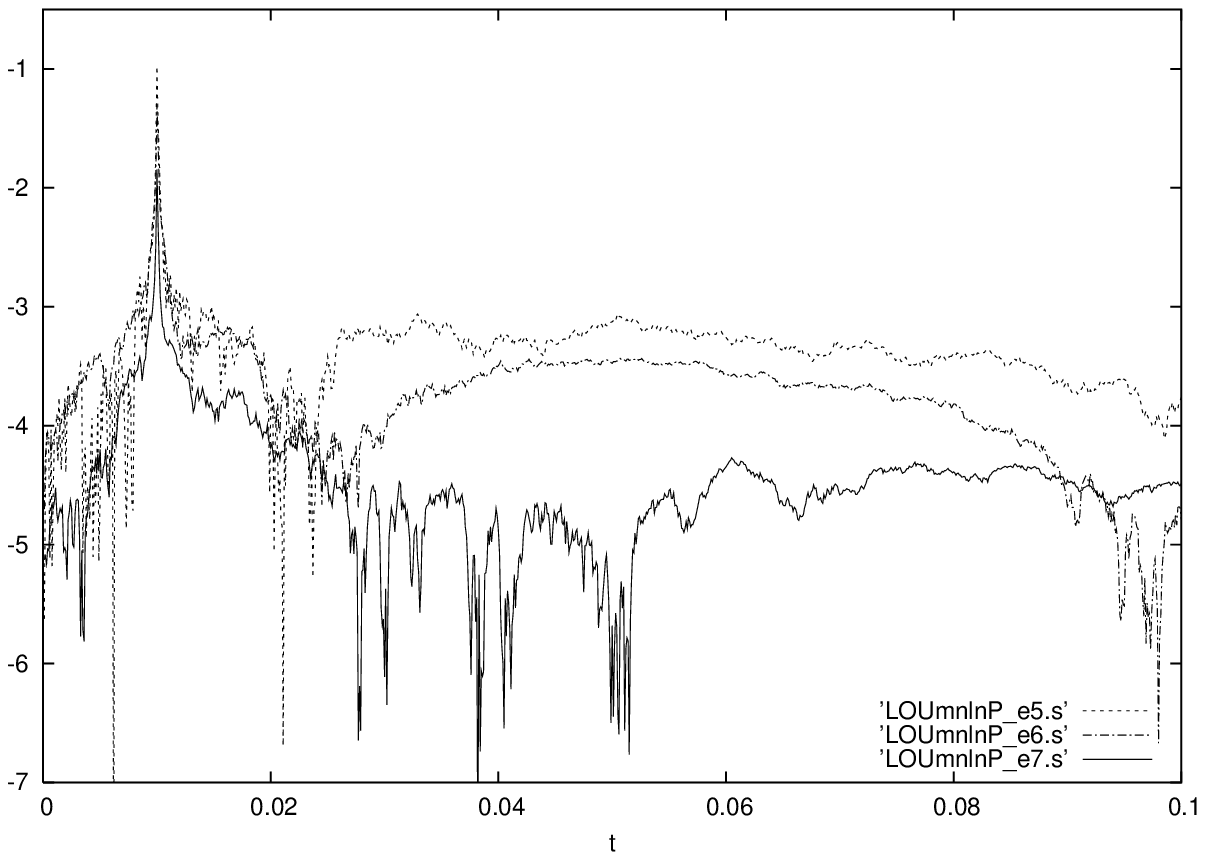,width=3in,height=2in}}
      }
    \caption{Error in mean of $\ln P_t$ for Log-Ornstein-Uhlenbeck model.}
    \label{LOU1}
  \end{center}
\end{figure}

In Fig. \ref{LOU1} we plot the log base ten relative error in $\overline{\ln P_t}$ for ANISE in (a) and SDE9 in (b). Errors are shown
for averages over $10^5$ (dashed curve), $10^6$ (dot-dashed curve) and $10^7$ (solid curve) trajectories. In all cases a spike
in error is seen near the time $t=.01$ where the exact $\overline{\ln P_t}$ passes through zero. The absolute error is small
and so the spike in relative error indicated in the 
plots is essentially fictitious and convergence is in fact good at all times for both methods.

\begin{figure}[h]
  \begin{center}
    \mbox{
      \subfigure[Error in $\overline{V_t}$ vs. $t$ for ANISE.]{\epsfig{file=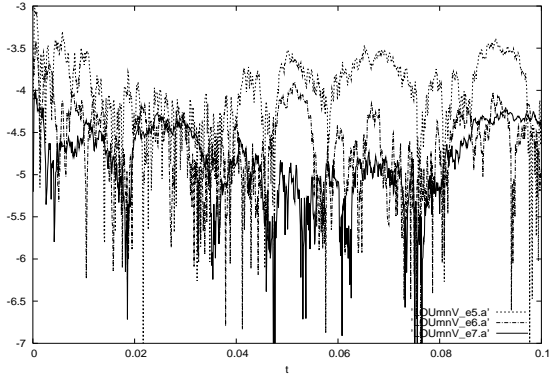,width=3in,height=2in}} \quad
      \subfigure[Error in $\overline{V_t}$ vs. $t$ for SDE9.]{\epsfig{file=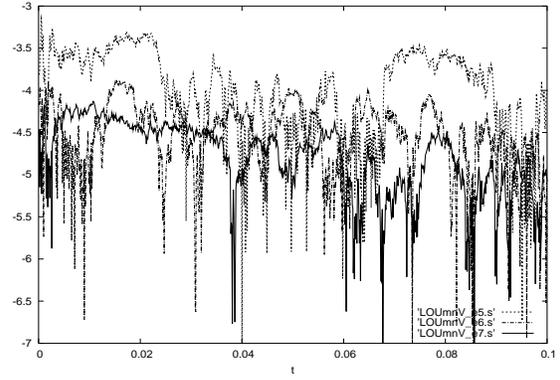,width=3in,height=2in}}
      }
    \mbox{
      \subfigure[Error in ${\rm var}(V_t)$ vs. $t$ for ANISE.]{\epsfig{file=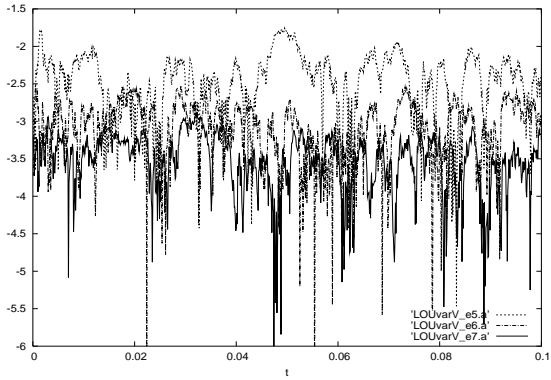,width=3in,height=2in}} \quad
      \subfigure[Error in ${\rm var}(V_t)$ vs. $t$ for SDE9.]{\epsfig{file=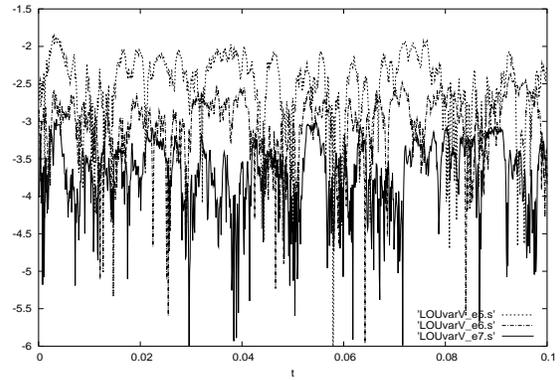,width=3in,height=2in}} 
      }
    \caption{Error in mean and variance of $V_t$ for Log-Ornstein-Uhlenbeck model.}
    \label{LOU2}
  \end{center}
\end{figure}

Figure \ref{LOU2} plots the mean relative error in the mean and variance of the volatility for ANISE ((a) and (c), respectively)
and SDE9 ((b) and (d), respectively). The error in the variance is larger than that in the mean. Here we see good convergence for both methods at all times.

The cpu times for various numbers of trajectories are shown in Table \ref{table:T4}. ANISE takes about 5 s to compute 1000 trajectories.
\begin{table}[h]
\begin{center}
\begin{tabular}{|c|c|c|c|}
\hline
~ \# Trajectories ~ &~~ ANISE CPU Time ~~&~~ SDE9 CPU Time ~~&~~ CPU Time Ratio SDE9/ANISE ~~ \\ \hline
~ $10^3$  & 0.53E+01 & 0.13E+02 & 2.46 \\ ~ $10^4$ & 0.52E+02 & 0.12E+03 & 2.38 \\~ $10^5$ & 0.52E+03 & 0.12E+04 & 2.29 \\~ $10^6$ & 0.52E+04 & 0.12E+05 & 2.28 \\ ~ $10^7$ & 0.52E+05 & 0.12E+06 & 2.29 \\
\hline
\end{tabular}
\end{center}
\caption{CPU times for Log-Ornstein-Uhlenbeck model in seconds.}
\label{table:T4}
\end{table}
The ratio of cpu time of SDE9 to ANISE is again larger than two. The ratio is roughly independent of the number
of trajectories.

\subsection{Affine Two Volatility Factor Model}

The equations for the log-price and volatilities of an affine two volatility model\cite{DK,Chern} are
\begin{eqnarray}
dP_t&=&\mu dt +\sqrt{\xi_0+\xi_1 V_{1t}+\xi_2V_{2t}}~dW_{3t}\\
dV_{1t}&=&(\alpha_{10}+\alpha_{11}V_{1t})dt+\sqrt{V_{1t}}~dW_{1t}\\
dV_{2t}&=&(\alpha_{20}+\alpha_{21}V_{2t})dt+\sqrt{V_{2t}}~dW_{2t}
\end{eqnarray}
and so we now have three equations and three Wiener processes.

The derivatives required by the numerical methods are given in Table \ref{table:D5}. A time step of $dt=.001$ was employed and the
equations were integrated to 0.5.
\begin{table}[h]
\begin{center}
\begin{tabular}{c|cccc}
~ $X_t$ ~ & ~ $\frac{\partial X_t}{\partial t}$ ~ & ~ $\frac{\partial X_t}{\partial W_{1t}}$ ~ & ~ $\frac{\partial X_t}{\partial W_{2t}}$ ~ & ~ $\frac{\partial X_t}{\partial W_{3t}}$ ~ \\ \hline 
 $P_t$ & ~~$\mu$~~ & 0 & 0 & $\sqrt{\xi_0-\xi_1 V_{1t}-\xi_2 V_{2t}}$\\ $V_{1t}$ & $\alpha_{10}+\alpha_{11}V_{1t}-\frac{1}{4}$ & $\sqrt{V_{1t}}$ & 0 & 0 \\ $V_{2t}$ & $\alpha_{20}+\alpha_{21}V_{2t}-\frac{1}{4}$  & 0 & $\sqrt{V_{2t}}$ & 0 \\ 
\end{tabular}
\end{center}
\caption{ Derivatives for Affine Two Volatility Factor Model. }
\label{table:D5}
\end{table}

We examined quantities $\overline{P_t}$, ${\rm var}(P_t)$, $\overline{V_{1t}}$, and ${\rm var}(V_{1t})$
which have exact solutions given by
\begin{eqnarray}
\overline{ P_t}&=&P_0+\mu t\\
{\rm var}(P_t)&=&[\xi_0-\frac{\xi_1\alpha_{10}}{\alpha_{11}}-\frac{\xi_2\alpha_{20}}{\alpha_{21}}]t+\frac{\xi_1}{\alpha_{11}}(V_{10}+\frac{\alpha_{10}}{\alpha_{11}})(e^{\alpha_{11}t}-1)\nonumber \\
&+&\frac{\xi_2}{\alpha_{21}}(V_{20}+\frac{\alpha_{20}}{\alpha_{21}})(e^{\alpha_{21}t}-1)\\
\overline{V_{1t}}&=&V_{10}e^{\alpha_{11}t}+\frac{\alpha_{10}}{\alpha_{11}}(e^{\alpha_{11}t}-1)\\
{\rm var}(V_{1t})&=&\frac{\alpha_{10}}{2\alpha_{11}^2}(e^{\alpha_{11}t}-1)^2+\frac{V_{10}}{\alpha_{11}}(e^{2\alpha_{11}t}-e^{\alpha_{11}t}).
\end{eqnarray}
The parameters were set as $\xi_0=.01$, $\xi_1=.1258$, $\xi_2=.0344$, $\mu=.02$, $\alpha_{10}=.2894$, $\alpha_{11}=17.4321$, $\alpha_{20}=.0602$, $\alpha_{21}=13.6036$, $P_0=1$, $V_{10}=.2$, and $V_{20}=.2$.
Note that on average $V_{1t}$ increases exponentially with a large exponent, and so the noises in the equation
for the price $P_t$ are strongly weighted.

\begin{figure}[h]
  \begin{center}
    \mbox{
      \subfigure[Error in $\overline{P_t}$ vs. $t$ for ANISE.]{\epsfig{file=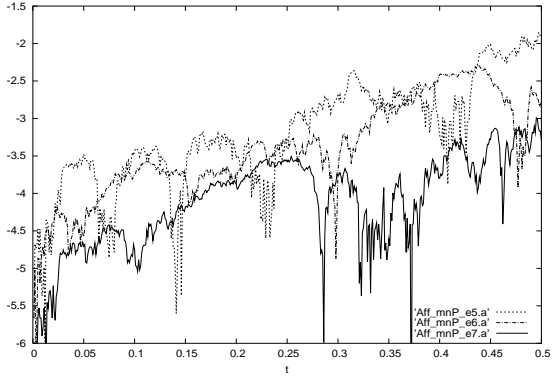,width=3in,height=2in}} \quad
      \subfigure[Error in $\overline{P_t}$ vs. $t$ for SDE9.]{\epsfig{file=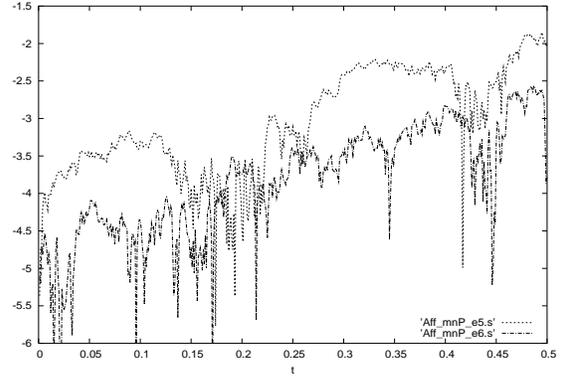,width=3in,height=2in}}
      }
    \mbox{
      \subfigure[Error in ${\rm var}(P_t)$ vs. $t$ for ANISE.]{\epsfig{file=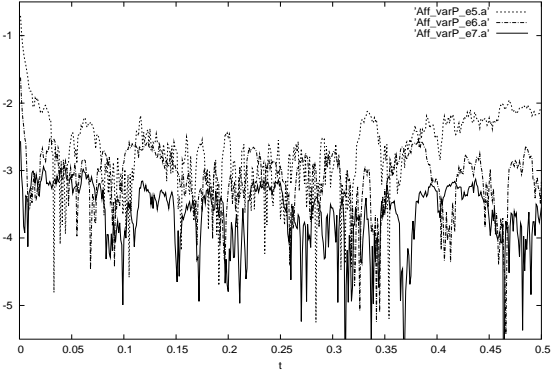,width=3in,height=2in}} \quad
      \subfigure[Error in ${\rm var}(P_t)$ vs. $t$ for SDE9.]{\epsfig{file=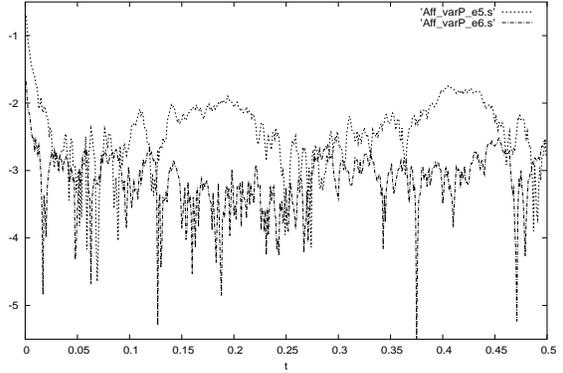,width=3in,height=2in}} 
      }
    \caption{Error in mean and variance of $P_t$ for Affine model.}
    \label{Aff1}
  \end{center}
\end{figure}

In Fig. \ref{Aff1} we show convergence via the log base ten relative error in the mean and variance of the price for ANISE ((a) and (c), respectively)
and SDE9 (b) and (d), respectively). For ANISE the plots show three curves corresponding to runs with averages over $10^5$ (dashed curve), $10^6$ (dot-dashed curve),
and $10^7$ (solid curve) trajectories. For SDE9 the plots show just two curves corresponding to runs with averages over $10^5$ (dashed curve) and $10^6$ (dot-dashed curve) trajectories. In both cases good convergence is observed toward the exact solution. The error in the variance is larger than that
in the mean. As we discuss below the relative cpu time
for SDE9 is much larger than for previous problems. Indeed, the run with $10^7$ trajectories did not finish and so does not appear in the figures.

\begin{figure}[htbp]
  \begin{center}
    \mbox{
      \subfigure[Error in $\overline{V_{1t}}$ vs. $t$ for ANISE.]{\epsfig{file=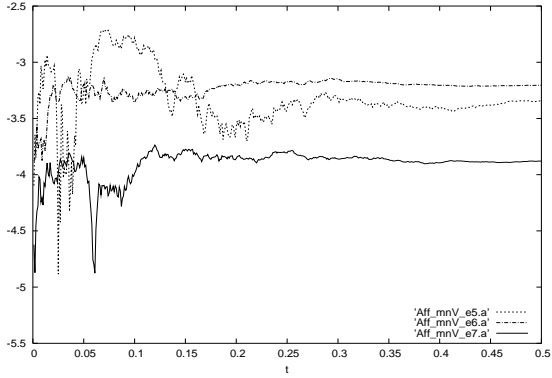,width=3in,height=2in}} \quad
      \subfigure[Error in $\overline{V_{1t}}$ vs. $t$ for SDE9.]{\epsfig{file=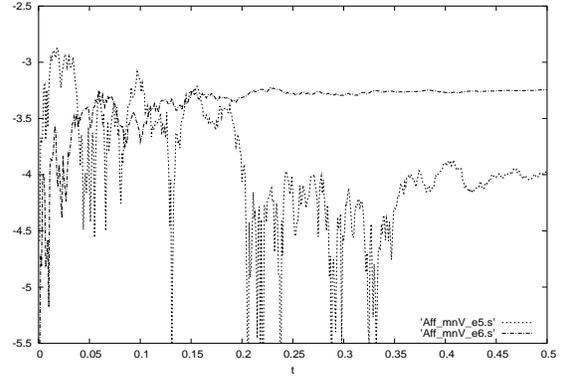,width=3in,height=2in}}
      }
    \mbox{
      \subfigure[Error in ${\rm var}(V_{1t})$ vs. $t$ for ANISE.]{\epsfig{file=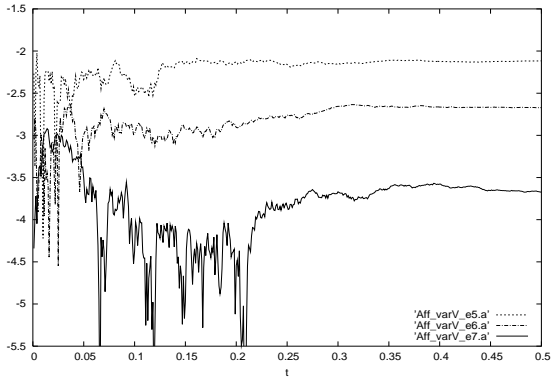,width=3in,height=2in}} \quad
      \subfigure[Error in ${\rm var}(V_{1t})$ vs. $t$ for SDE9.]{\epsfig{file=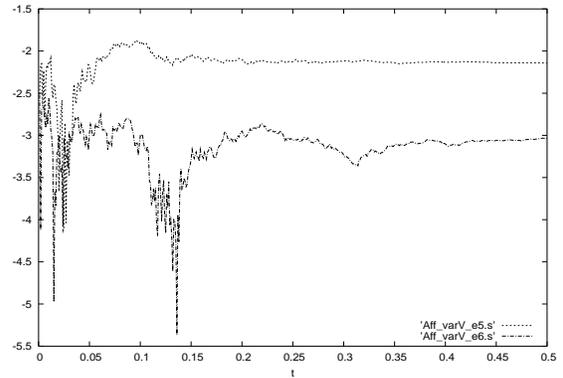,width=3in,height=2in}} 
      }
    \caption{Error in mean and variance of $V_{1t}$ for Affine model.}
    \label{Aff2}
  \end{center}
\end{figure}

Figure \ref{Aff2} plots the errors in $\overline{V_{1t}}$ and ${\rm var}(V_{1t})$ against time for ANISE ((a) and (c), respectively) and SDE9 (b) and (d), respectively) for the same numbers of trajectories as in the previous figure. Error in the variance is larger than that in the mean. Good convergence is again observed for both methods.

The cpu times for various numbers of trajectories are shown in Table \ref{table:T5}. ANISE takes about 3.5 s to compute 1000 trajectories. In spite of the fact that good convergence was 
observed for the SDE9 method its computation times show a large jump from the $10^4$ calculation to the $10^5$ calculation. For
$10^5$ and $10^6$ ANISE is several hundred times faster than SDE9. Rare
trajectories with difficult stochastic paths appear to be responsible for the poor performance of SDE9. 
\begin{table}[h]
\begin{center}
\begin{tabular}{|c|c|c|c|}
\hline
~ \# Trajectories ~ &~~ ANISE CPU Time ~~&~~ SDE9 CPU Time ~~&~~ CPU Time Ratio SDE9/ANISE ~~ \\ \hline
~ $10^3$ & 0.34E+01 & 0.10E+02 & 2.96 \\ ~ $10^4$ & 0.34E+02 & 0.92E+02 & 2.74 \\~ $10^5$ & 0.34E+03 & 0.88E+05 & 261.61 \\~ $10^6$ & 0.34E+04 & 0.13E+07 & 386.63 \\ ~ $10^7$ & 0.34E+05 & NA & NA \\
\hline
\end{tabular}
\end{center}
\caption{CPU times for Affine model in seconds.}
\label{table:T5}
\end{table}

\subsection{Log Linear Two Volatility Factor Model Without Feedback}

The price and volatilities obey\cite{Chern}
\begin{eqnarray}
dP_t&=&\{(\alpha_{10}+\alpha_{12}V_{2t})dt+e^{\beta_{10}+\beta_{13}V_{3t}}[\sqrt{1-\psi_{13}^2}dW_{1t}+\psi_{13}dW_{3t}]\}P_t\\
dV_{2t}&=&\alpha_{22}V_{2t}dt+dW_{2t}\\
dV_{3t}&=&\alpha_{33}V_{3t}dt+dW_{3t}
\end{eqnarray}
and so we have three equations and three Wiener processes.

The derivatives needed by the numerical methods are given in Table \ref{table:D6}. A time step of $10^{-4}$ was used and the equations
were integrated to 0.1.
\begin{table}[h]
\begin{center}
\begin{tabular}{c|cccc}
~ $X_t$ ~ & ~ $\frac{\partial X_t}{\partial t}$ ~ & ~ $\frac{\partial X_t}{\partial W_{1t}}$ ~ & ~ $\frac{\partial X_t}{\partial W_{2t}}$ ~ & ~ $\frac{\partial X_t}{\partial W_{3t}}$ ~ \\ \hline 
 $P_t$ & ~~~$[\alpha_{10}+\alpha_{12}V_{2t}-\frac{1}{2} F_t^2 -\frac{1}{2}\beta_{13}\psi_{13}F_t]P_t$ $~~~~$& $\sqrt{1-(\psi_{13})^2}~F_tP_t$ & 0 & $\psi_{13}F_tP_t$ \\
$V_{2t}$ & $\alpha_{22}V_{2t}$ & 0 & 1 & 0 \\ 
$V_{3t}$ & $\alpha_{33}V_{3t}$  & 0 & 0 & 1 \\ 
\end{tabular}
\end{center}
\caption{ Derivatives for Log Linear model without feedback. Here $F_t=e^{\beta_{10}+\beta_{13}V_{3t}}$. }
\label{table:D6}
\end{table}

We calculated $\overline{\ln P_t}$, $\overline{V_{2t}}$, $\overline{V_{3t}}$, and ${\rm var}(V_{3t})$,
some of which have known exact solutions
\begin{eqnarray}
\overline{V_{2t}}&=&V_{20}e^{\alpha_{22}t}\\
\overline{V_{3t}}&=&V_{30}e^{\alpha_{33}t}\\
{\rm var}(V_{3t})&=&\frac{1}{2\alpha_{33}}(e^{2\alpha_{33}t}-1).
\end{eqnarray}
Once again we had to solve an ODE
\begin{eqnarray}
\frac{d\overline{\ln P_t}}{dt}=\alpha_{10}+\alpha_{12}V_{20}e^{\alpha_{22}t}-\frac{1}{2}\exp\{2\beta_{10}+2\beta_{13}V_{30}e^{\alpha_{33}t}+\frac{\beta_{13}^2}{\alpha_{33}}(e^{2\alpha_{33}t}-1)\}
\end{eqnarray}
numerically to find $\overline{\ln P_t}$. This was again accomplished using a Runge-Kutta algorithm for ODEs\cite{Hair}.

The parameters were set as $\alpha_{10}=.0337$, $\alpha_{12}=.4820$, $\alpha_{22}=1.0043$,
$\alpha_{33}=.0291$, $\beta_{10}=1.0294$, $\beta_{13}=.0261$, 
$\psi_{13}=.3285$, $P_0=1$, $V_{20}=.1$, $V_{30}=.05$.

\begin{figure}[h]
  \begin{center}
    \mbox{
      \subfigure[$\overline{\ln P_t}$ vs. $t$ for ANISE]{\epsfig{file=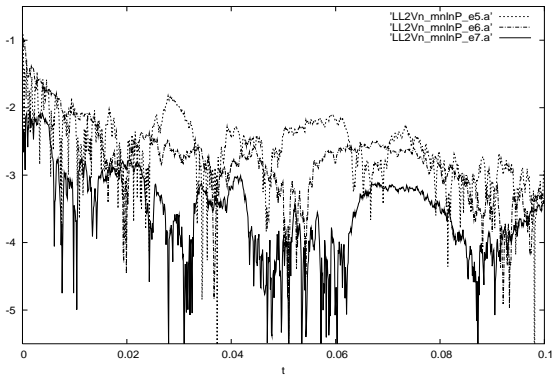,width=3in,height=2in}} \quad
      \subfigure[$\overline{\ln P_t}$ vs. $t$ for SDE9]{\epsfig{file=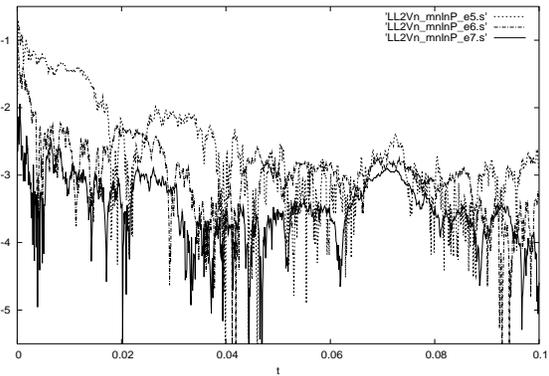,width=3in,height=2in}}
      }
    \mbox{
      \subfigure[$\overline{V_{2t}}$ vs. $t$ for ANISE]{\epsfig{file=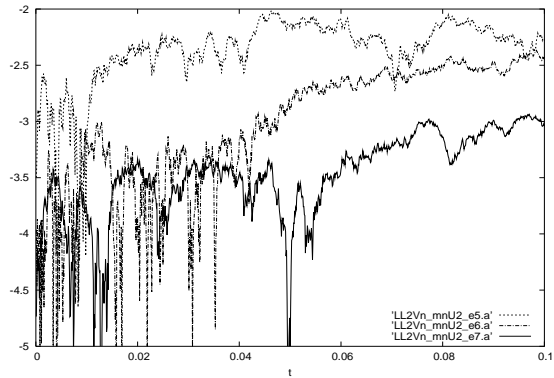,width=3in,height=2in}} \quad
      \subfigure[$\overline{V_{2t}}$ vs. $t$ for SDE9]{\epsfig{file=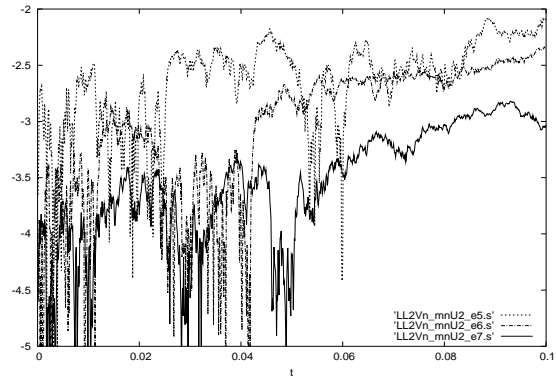,width=3in,height=2in}} 
      }
    \caption{Means of $\ln P_t$ and $V_{2t}$ for Log Linear model without feedback}
    \label{LL2Vn1}
  \end{center}
\end{figure}

In Fig. \ref{LL2Vn1} we plot the log base ten relative error in $\overline{\ln P_t}$ and $\overline{V_{2t}}$ for ANISE ((a) and (c), respectively)
and SDE9 ((b) and (d), respectively). In all cases the dashed curve represents an average over $10^5$ trajectories while the dot-dashed and 
solid curves are for $10^6$ and $10^7$ trajectories, respectively. Good convergence is seen in all cases except near $t=0$
for $\overline{\ln P_t}$. The exact solution for $\overline{\ln P_t}$ vanishes at $t=0$ for our initial condition, and
poor relative accuracy is seen as a consequence. In fact the absolute accuracy is good at all times for $10^7$ trajectories. 

\begin{figure}[h]
  \begin{center}
    \mbox{
      \subfigure[$\overline{V_{3t}}$ vs. $t$ for ANISE]{\epsfig{file=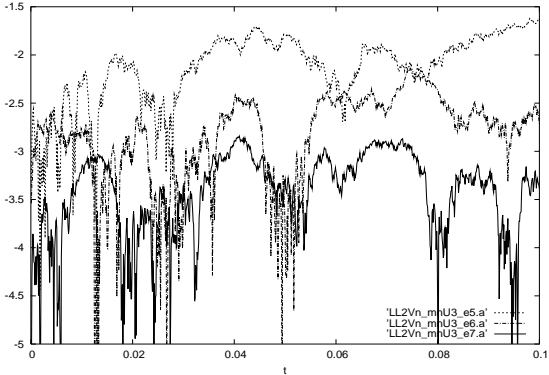,width=3in,height=2in}} \quad
      \subfigure[$\overline{V_{3t}}$ vs. $t$ for SDE9]{\epsfig{file=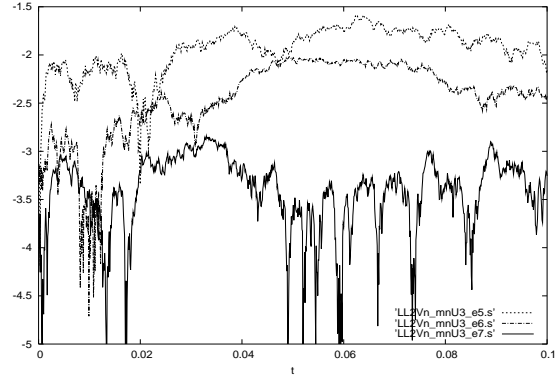,width=3in,height=2in}}
      }
    \mbox{
      \subfigure[${\rm var}(V_{3t})$ vs. $t$ for ANISE]{\epsfig{file=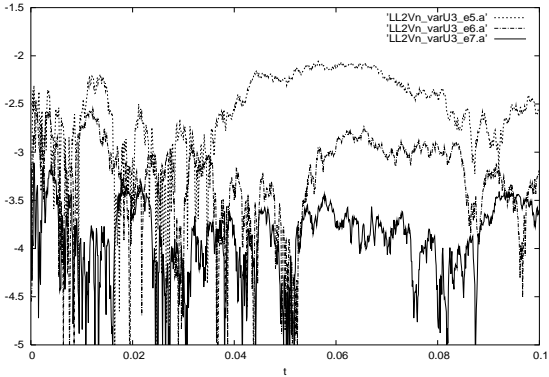,width=3in,height=2in}} \quad
      \subfigure[${\rm var}(V_{3t})$ vs. $t$ for SDE9]{\epsfig{file=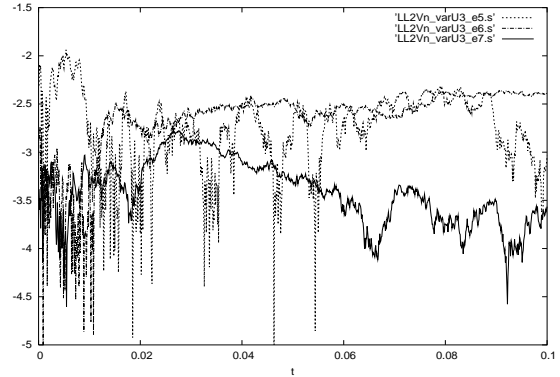,width=3in,height=2in}} 
      }
    \caption{Error in mean and variance of $V_{3t}$ for Log Linear model without feedback}
    \label{LL2Vn2}
  \end{center}
\end{figure}

Figure \ref{LL2Vn2} plots the errors in $\overline{V_{3t}}$ and ${\rm var}(V_{3t})$ against time for ANISE and SDE9. Once again, good convergence
is observed. Errors in the mean and variance are comparable.

The cpu times for various numbers of trajectories are shown in Table \ref{table:T6}. ANISE takes 6.5 s to compute 1000 trajectories. Once again the ratio of cpu time for SDE9 to that of ANISE
is a little greater than two and this number is independent of the number of trajectories.
\begin{table}[h]
\begin{center}
\begin{tabular}{|c|c|c|c|}
\hline
~ \# Trajectories ~ &~~ ANISE CPU Time ~~&~~ SDE9 CPU Time ~~&~~ CPU Time Ratio SDE9/ANISE ~~ \\ \hline
~ $10^3$  & 0.65E+01 & 0.14E+02 & 2.11 \\ ~ $10^4$ & 0.64E+02 & 0.14E+03 & 2.13 \\~ $10^5$ & 0.64E+03 & 0.14E+04 & 2.18 \\~ $10^6$ & 0.64E+04 & 0.14E+05 & 2.18 \\ ~ $10^7$ & 0.64E+05 & 0.14E+06 & 2.17 \\
\hline
\end{tabular}
\end{center}
\caption{CPU times for Log Linear model without feedback in seconds. }
\label{table:T6}
\end{table}

\subsection{Log Linear Two Volatility Factor Model With Feedback}

The equations for this model\cite{Chern} are
\begin{eqnarray}
dP_t&=&\{(\alpha_{10}+\alpha_{12}V_{2t})dt\nonumber \\
&+&e^{\beta_{10}+\beta_{13}V_{3t}+\beta_{14}V_{4t}}[\sqrt{1-\psi_{13}^2-\psi_{14}^2}dW_{1t}+\psi_{13}dW_{3t}+\psi_{14}dW_{4t}]\}P_t\\
dV_{2t}&=&\alpha_{22}V_{2t}dt+dW_{2t}\\
dV_{3t}&=&\alpha_{33}V_{3t}dt+(1+\beta_{33}V_{3t})dW_{3t}\\
dV_{4t}&=&\alpha_{44}V_{4t}dt+(1+\beta_{44}V_{4t})dW_{4t}.
\end{eqnarray}
In this case we have four equations and four Wiener processes.

The derivatives required by the numerical methods are given in Table \ref{table:D7}. A time step of $dt=10^{-4}$ was employed and the equations
were integrated to 0.1.
\begin{table}[h]
\begin{center}
\begin{tabular}{c|ccccc}
~ $X_t$ ~ & ~ $\frac{\partial X_t}{\partial t}$ ~ & ~ $\frac{\partial X_t}{\partial W_{1t}}$ ~ & ~ $\frac{\partial X_t}{\partial W_{2t}}$ ~ & ~ $\frac{\partial X_t}{\partial W_{3t}}$ ~ & ~ $\frac{\partial X_t}{\partial W_{4t}}$ \\ \hline 
 $P_t$ & ~~~$\{\alpha_{10}+\alpha_{12}V_{2t}-\frac{1}{2} F_t^2$~~~~& $\rho F_tP_t $ & 0 & ~~$\psi_{13}F_t P_t $~~ & ~~$\psi_{14}F_tP_t$~~\\

~&~$ -\frac{1}{2}[\psi_{13}\beta_{13}(1+\beta_{33}V_{3t})+\frac{1}{2}\psi_{14}\beta_{14}(1+\beta_{44}V_{4t})]F_t \}P_t     $~&~~~~&~&~& \\
$V_{2t}$ & $\alpha_{22}V_{2t}$ & 0 & 1 & 0 & 0\\ $V_{3t}$ & $\alpha_{33}V_{3t}-\frac{1}{2}\beta_{33}(1+\beta_{33}V_{3t})$  & 0 & 0 & $1+\beta_{33}V_{3t}$ & 0\\$V_{4t}$ & $\alpha_{44}V_{4t}-\frac{1}{2}\beta_{44}(1+\beta_{44}V_{4t})$  & 0 & 0 & 0 & $1+\beta_{44}V_{4t}$\\
\end{tabular}
\end{center}
\caption{ Derivatives for Log Linear model with feedback. Here $F_t=e^{\beta_{10}+\beta_{13}V_{3t}+\beta_{14}V_{4t}} $  and $\rho=\sqrt{1-\psi_{13}^2-\psi_{14}^2}$.}
\label{table:D7}
\end{table}

We examined quantities $\overline{\ln P_t}$, $\overline{V_{2t}}$, $\overline{V_{3t}}$, and ${\rm var}(V_{3t})$
some of which have exact solutions
\begin{eqnarray}
\overline{V_{2t}}&=&V_{20}e^{\alpha_{22}t}\\
\overline{V_{3t}}&=&V_{30}e^{\alpha_{33}t}\\
{\rm var}(V_{3t})&=&(V_{30})^2(e^{(2\alpha_{33}+\beta_{33}^2)t}-e^{2\alpha_{33}t})
+\frac{2\beta_{33}V_{30}}{\alpha_{33}+\beta_{33}^2}(e^{(2\alpha_{33}+\beta_{33}^2)t}-e^{\alpha_{33}t})\nonumber \\
&+&\frac{1}{2\alpha_{33}+\beta_{33}^2}(e^{(2\alpha_{33}+\beta_{33}^2)t}-1).
\end{eqnarray}
We obtained $\overline{\ln P_t}$ numerically by solving the ordinary differential equation
\begin{eqnarray}
\frac{d\overline{\ln P_t}}{dt}=\alpha_{10}+\alpha_{12}V_{20}e^{\alpha_{22}t}-\frac{e^{2\beta_{10}}}{2}\overline{e^{2\beta_{13}V_{3t}}}~\overline{e^{2\beta_{14}V_{4t}}}
\end{eqnarray}
using a variable-stepsize Runge-Kutta scheme\cite{Hair}. The averages $\overline{e^{2\beta_{1i}V_{it}}}$ 
for $i=3,4$ were obtained from the moments $\overline{(V_{it})^n}$ using $\overline{e^{xV_{it}}}
=\sum_{n=0}^{\infty} \frac{x^n}{n!}\overline{(V_{it})^n}$ (numerically truncated after $n=20$) and iteration using
\begin{eqnarray}
\overline{V_{it}}&=&V_{i0}e^{\alpha_{ii}t}\\
\overline{(V_{it})^2}&=&(V_{i0})^2e^{(2\alpha_{ii}+\beta_{ii}^2)t}
+\frac{2\beta_{ii}V_{i0}}{\alpha_{ii}+\beta_{ii}^2}(e^{(2\alpha_{ii}+\beta_{ii}^2)t}-e^{\alpha_{ii}t})+\frac{1}{2\alpha_{ii}+\beta_{ii}^2}(e^{(2\alpha_{ii}+\beta_{ii}^2)t}-1)\\
\overline{(V_{it})^n}&=&(V_{i0})^ne^{(n\alpha_{ii}+\frac{n(n-1)}{2}\beta_{ii}^2)t}
+n(n-1)\beta_{ii}\int_0^tdt' ~\overline{(V_{it'})^{n-1}}~e^{(n\alpha_{ii}+\frac{n(n-1)}{2}\beta_{ii}^2)(t-t')}\nonumber \\
&+&\frac{n(n-1)}{2}\int_0^tdt'~\overline{(V_{it'})^{n-2}}~e^{(n\alpha_{ii}+\frac{n(n-1)}{2}\beta_{ii}^2)(t-t')},~~~~~~{\rm for}~~~~~n=3,4,\dots
\end{eqnarray}
which are also readily obtained using an ODE code.

The parameters were set as $\alpha_{10}=.0279$, $\alpha_{12}=.7281$, $\alpha_{22}=5.9997$,
$\alpha_{33}=.1227$, $\alpha_{44}=8.2119$, $\beta_{10}=.0486$, $\beta_{13}=.0695$, 
$\beta_{14}=.3130$, $\beta_{33}=.3672$, $\beta_{44}=.3655$, $\psi_{13}=.1077$, $\psi_{14}=.0564$, with initial conditions $P_0=1$, $V_{20}=.1$, $V_{30}=.05$, $V_{40}=.2$.

\begin{figure}[h]
  \begin{center}
    \mbox{
      \subfigure[$\overline{\ln P_t}$ vs. $t$ for ANISE]{\epsfig{file=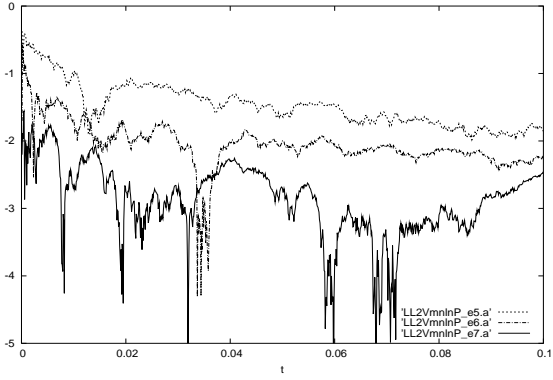,width=3in,height=2in}} \quad
      \subfigure[$\overline{\ln P_t}$ vs. $t$ for SDE9]{\epsfig{file=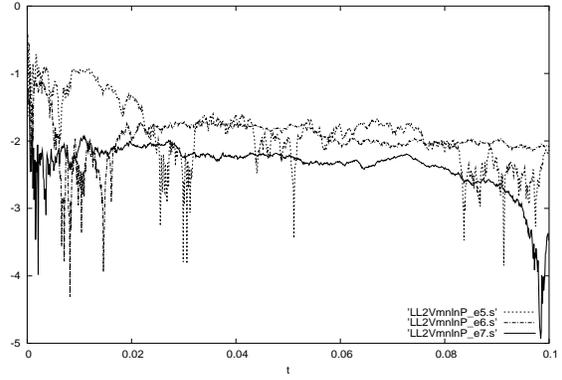,width=3in,height=2in}}
      }
    \mbox{
      \subfigure[$\overline{V_{2t}}$ vs. $t$ for ANISE]{\epsfig{file=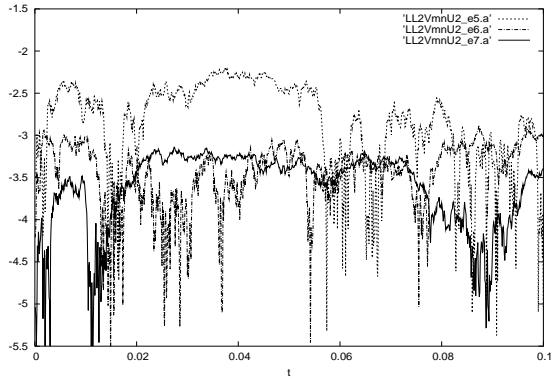,width=3in,height=2in}} \quad
      \subfigure[$\overline{V_{2t}}$ vs. $t$ for SDE9]{\epsfig{file=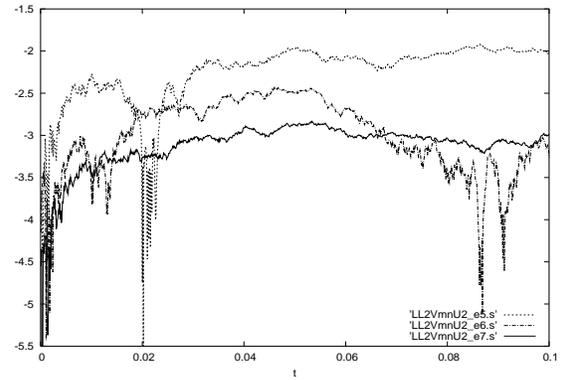,width=3in,height=2in}} 
      }
    \caption{Means of $\ln P_t$ and $V_{2t}$ for Log Linear model with feedback}
    \label{LL2V1}
  \end{center}
\end{figure}

In Fig. \ref{LL2V1} we plot the log base ten relative accuracy of $\overline{\ln P_t}$ and $\overline{V_{2t}}$ against time
for ANISE ((a) and (c), respectively) and SDE9 ((b) and (d), respectively). The dashed curve represents an average over
$10^5$ trajectories, while the dot-dashed and solid curves represent calculations with $10^6$ and $10^7$ trajectories, 
respectively. In all cases convergence is good except for $\overline{\ln P_t}$ in the vicinity of zero where the exact
solution vanishes and the relative accuracy becomes poorly defined. 

\begin{figure}[h]
  \begin{center}
    \mbox{
      \subfigure[$\overline{V_{3t}}$ vs. $t$ for ANISE]{\epsfig{file=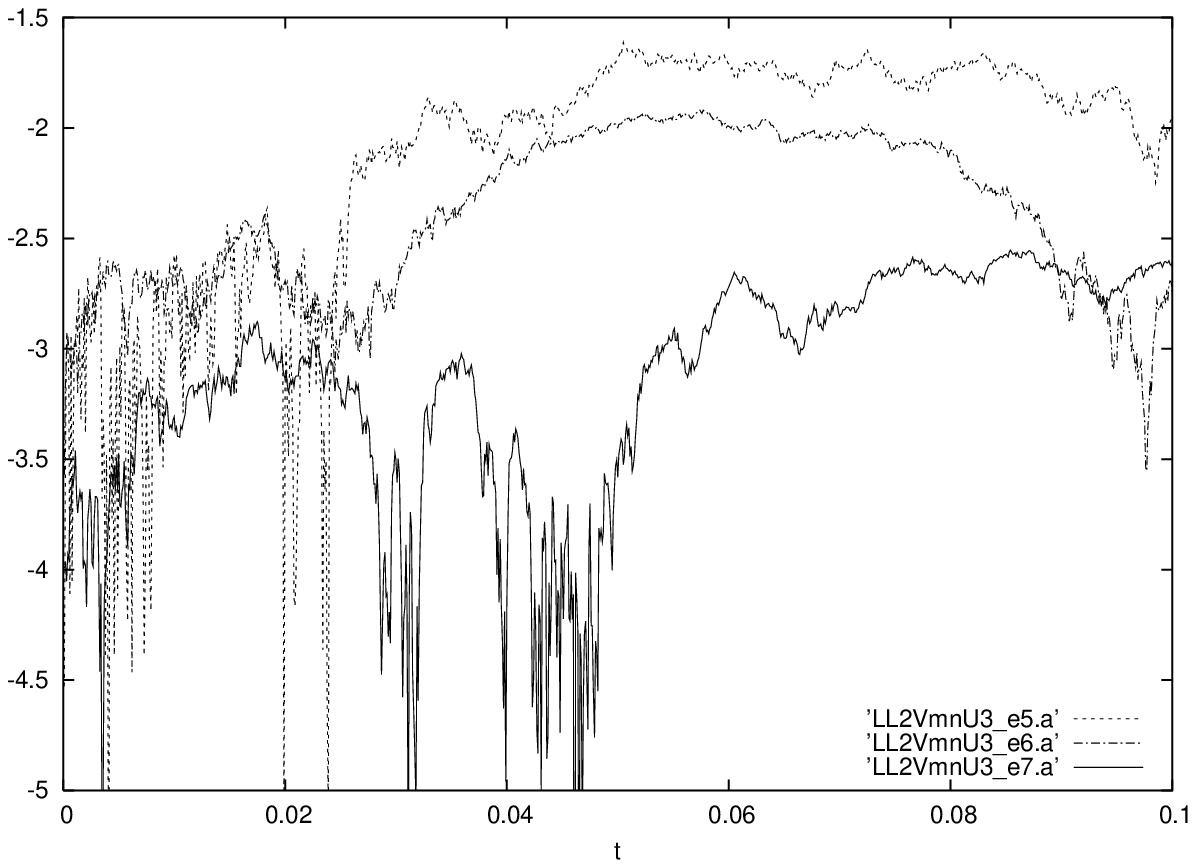,width=3in,height=2in}} \quad
      \subfigure[$\overline{V_{3t}}$ vs. $t$ for SDE9]{\epsfig{file=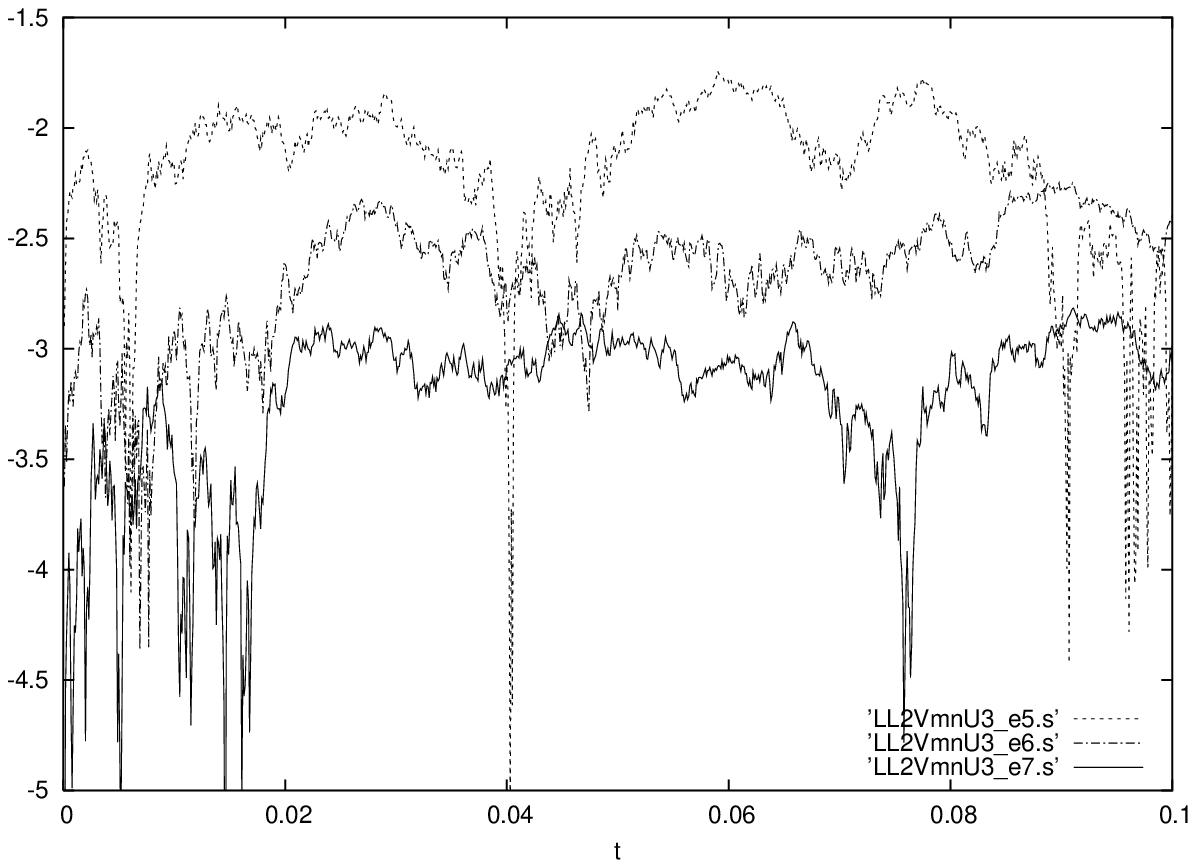,width=3in,height=2in}}
      }
    \mbox{
      \subfigure[${\rm var}(V_{3t})$ vs. $t$ for ANISE]{\epsfig{file=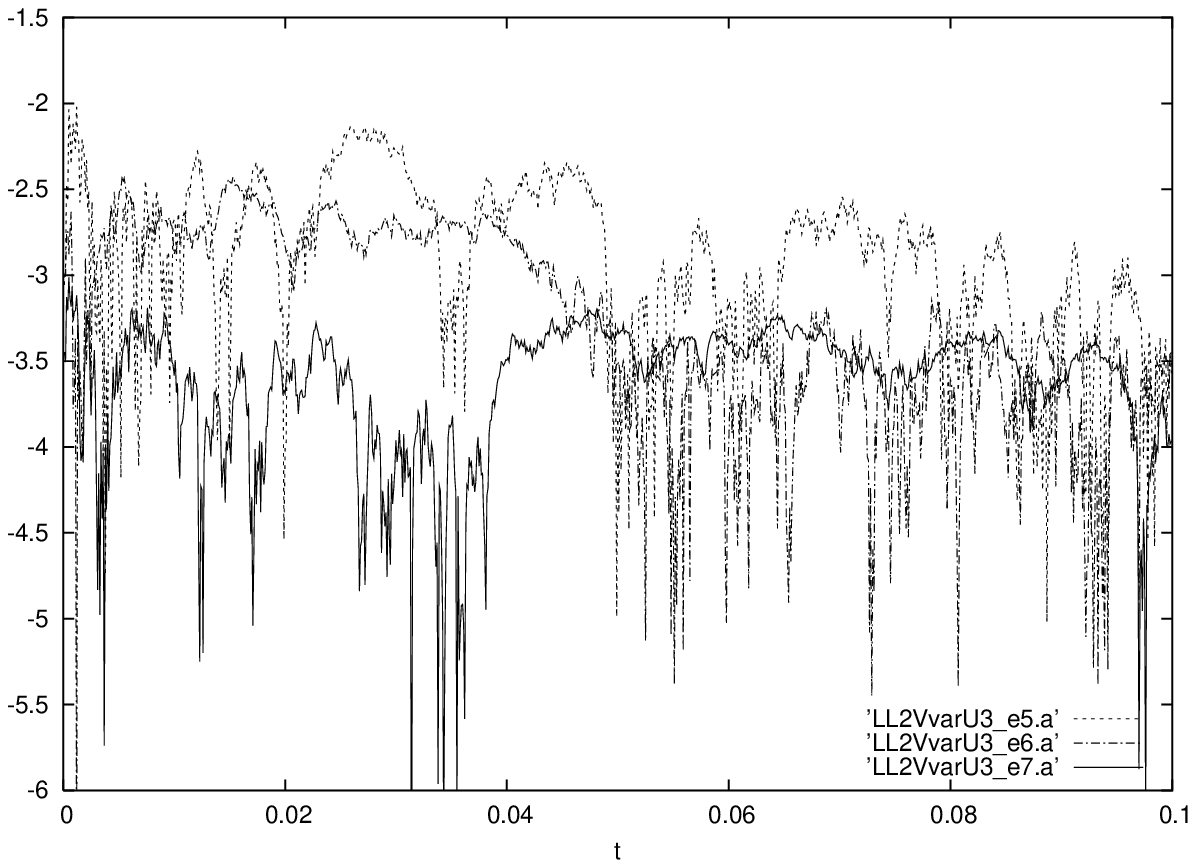,width=3in,height=2in}} \quad
      \subfigure[${\rm var}(V_{3t})$ vs. $t$ for SDE9]{\epsfig{file=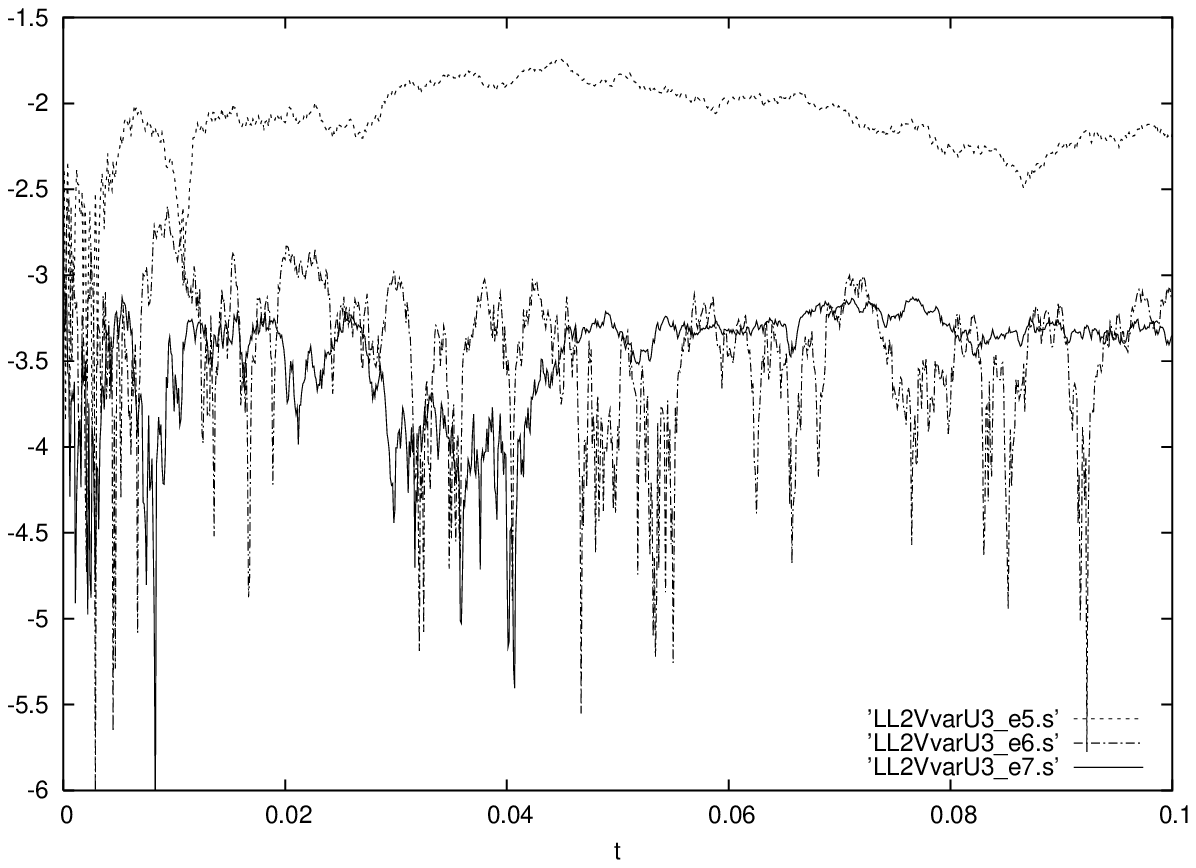,width=3in,height=2in}} 
      }
    \caption{Error in mean and variance of $V_{3t}$ for Log Linear model with feedback}
    \label{LL2V2}
  \end{center}
\end{figure}

Figure \ref{LL2V2} plots the error in $\overline{V_{3t}}$ and ${\rm var}(V_{3t})$ for ANISE and SDE9. Good convergence 
is observed in all cases. Errors in the mean are greater than those in the variance.

The cpu times for various numbers of trajectories are given in Table \ref{table:T7}. ANISE takes 7.5 s to compute 1000 trajectories. Once again ANISE is about twice as fast as SDE9.
\begin{table}[h]
\begin{center}
\begin{tabular}{|c|c|c|c|}
\hline
~ \# Trajectories ~ &~~ ANISE CPU Time ~~&~~ SDE9 CPU Time ~~&~~ CPU Time Ratio SDE9/ANISE ~~ \\ \hline
~ $10^3$ & 0.74E+01 & 0.15E+02 & 2.04 \\ ~ $10^4$ & 0.75E+02 & 0.15E+03 & 2.06 \\~ $10^5$ & 0.74E+03 & 0.16E+04 & 2.08 \\~ $10^6$ & 0.75E+04 & 0.15E+05 & 2.08 \\ ~ $10^7$ & 0.75E+05 & 0.15E+06 & 2.08 \\
\hline
\end{tabular}
\end{center}
\caption{CPU times for Log Linear model with feedback in seconds.}
\label{table:T7}
\end{table}

\section{Accuracy for Individual Trajectories}

Here we again request a relative accuracy of $10^{-12}$ and determine what accuracy is in fact obtained on average for individual trajectories. While it is unlikely that results of this high precision would be required in actual financial applications, it is worth exploring this issue for a few problems where exact solutions of the SODEs are known. We find that the calculations are not very sensitive to the requested tolerance, and accuracies of $10^{-12}$ are sometimes achieved even when the requested tolerance is only $10^{-6}$. The calculations are
also insensitive to the stepsize.

For each realization of the observable we thus calculate an exact solution $X_t$ and an approximate solution $X_t^{approx.}$ from 
which we compute
the log base ten relative error
\begin{equation}
\log_{10}[\frac{|X_t-X_t^{approx.}|}{{\rm max}\{|X_t|,|X_t^{approx.}|\}}]. 
\end{equation}
We plot the average of this quantity against time t for each model. The exact solutions involve some difficult integrals which are also computed using the numerical method, so our tests are essentially self-consistency checks. 

For both models we have requested large time steps and integrated to very long times in order to make the calculation somewhat challenging.
The errors shown are computed as time averages over short intervals since there are high frequency fluctuations in the data which
make identification of the line types in the figures difficult.

\subsection{Vasicek interest rate model}

The SODE for this model is 
\begin{equation}
 dV_t=c(\mu-V_t)dt+\sigma dW_t, 
\end{equation}
which has the solution
\begin{equation}
 V_t=V_0 e^{-ct}+ \mu(1-e^{-ct})+\sigma e^{-ct} \int_0^t e^{cs}dW_s.
\end{equation}
The derivatives needed by the numerical methods are given in Table \ref{table:indiv}.
\begin{table}[h]
\begin{center}
%\begin{tabular}{|c|c|c|c|c|c|}
\begin{tabular}{c|cc}
~ Model ~ & ~ $\frac{\partial V_t}{\partial t}$ ~ & ~ $\frac{\partial V_t}{\partial W_t}$ ~  \\ \hline 
 Vasicek & $c(\mu-V_t)$ & $\sigma$  \\ CEV & ~~ $\kappa(\theta-V_t)-\frac{1}{2}\alpha^2 V_t$ ~~ & $\alpha V_t$  \\ 
\end{tabular}
\end{center}
\caption{ equation array for Vasicek and CEV Models }
\label{table:indiv}
\end{table}

We set the parameters to $c$ = .05, $\mu$ = .09, $\sigma$ = .03, and $V_0$ = .08. We set the time step to $dt$ = 2.4 and integrated to 12000. This is of course a very long dynamics. We plot the average relative error in Fig. 15 (a) for ANISE (solid curve) and SDE9 (dot-dashed curve). Both ANISE and SDE9 return results consistent with the requested tolerance. SDE9 returns a greater relative tolerance than that requested. 

The cpu times are compared in Table \ref{table:models}. Here we see that SDE9 also runs somewhat faster than ANISE for this problem.

\begin{table}[h]
\begin{center}
\begin{tabular}{|c|c|c|c|}
\hline
~~ Model ~~ &~~~ ANISE ~~~&~~~ SDE9 ~~~&~~ Ratio SDE9/ANISE ~~ \\ \hline
~ Vasicek ~& .79E+05 & .57E+05 & 0.72 \\~ CEV ~& .37E+05 & .65E+05 & 1.75 \\
\hline
\end{tabular}
\end{center}
\caption{CPU times for tol = $10^{-12}$ and 1 million trajectories.}
\label{table:models}
\end{table}

\subsection{Mean-reverting CEV model}

Here the SODE is of the form\cite{Cox}
\begin{equation}
 dV_t=\kappa(\theta-V_t)dt+\alpha V_t dW_t 
\end{equation}
which has the exact solution
\begin{equation}
 V_t=\exp\{-(\kappa+\frac{1}{2}\alpha^2)t+\alpha W_t\}~[V_0+\kappa \theta \int_0^t ds~\exp\{(\kappa+\frac{1}{2}\alpha^2)s-\alpha W_s\}].
\end{equation}
The derivatives needed by the numerical methods are given in Table \ref{table:indiv}.

The parameters were chosen as  $\kappa$ = .05, $\theta$ = .09, $\alpha$ = .1, and $V_0$ = .08. We set the time step to $dt$ = 5 and integrated to 10000. The average relative error is plotted in Fig. 15 (b) for ANISE (solid curve) and SDE9 (dot-dashed curve). Both ANISE and SDE9 return results consistent with the requested tolerance. Once again SDE9 returns a better relative tolerance than that requested.

\begin{figure}[h]
  \begin{center}
    \mbox{
      \subfigure[ANISE (solid) and SDE9 (dot-dashed) for Vasicek.]{\epsfig{file=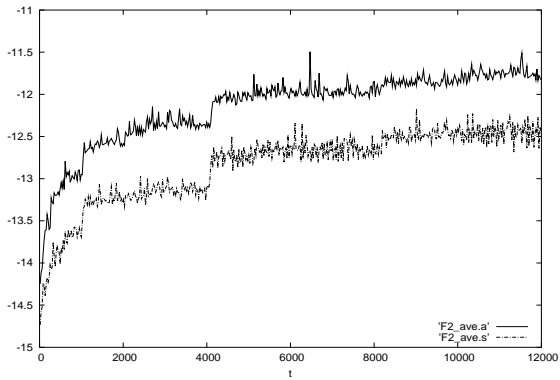,width=3in,height=2in}} \quad
      \subfigure[ANISE (solid) and SDE9 (dot-dashed) for CEV.]{\epsfig{file=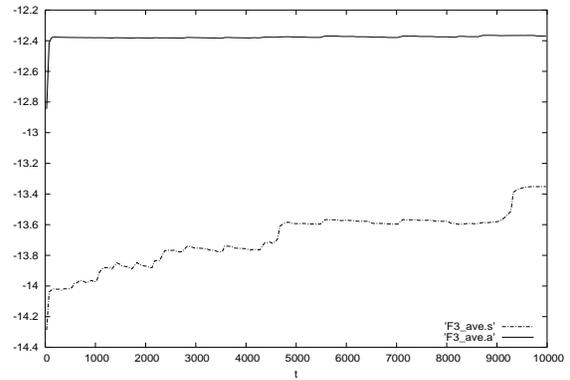,width=3in,height=2in}}
      }
    \caption{Average error in volatility $V_t$ for individual trajectories.}
    \label{F2&F3}
  \end{center}
\end{figure} 

Table \ref{table:models} contains the cpu times for the two methods. SDE9 takes 75 \% longer than ANISE.

\section{Conclusions}

Good convergence is obtainable using both ANISE and SDE9 for all the problems considered. In most cases ANISE runs roughly twice as fast. For the Vasicek model in section III SDE9 performed 40 \% faster than ANISE. ANISE performed several hundreds of times faster
than SDE9 for the Affine model in section II. 

In addition to our study of convergence, we examined the accuracy of individual trajectories for a given 
requested relative accuracy. We found that both methods returned trajectories with relative accuracies
consistent with the accuracy requested, even for very long integration times.

Both algorithms appear to be sufficiently accurate for the models considered. ANISE performed better overall. 
The two methods appear capable of handling larger systems of equations with more Wiener processes, and could
therefore prove to be valuable computational tools for further research in finance.

\section*{ACKNOWLEDGMENTS}
J.W. acknowledges the support of the Natural Sciences and Engineering Research Council of Canada.


\begin{thebibliography}{99}

\bibitem{Pearson} Pearson N.D.and Sun T.-S. 1994 Exploiting the conditional density in estimating the term structure: an application to the Cox, Ingersoll, and Ross model {\em J. Finance} 49, 1279-1304

\bibitem{CIR} Cox J.C., Ingersoll J.E. and Ross S.A. 1985 An intertemporal general equilibrium model of asset prices {\em Econometrica} 53, 363-384

\bibitem{Cox} Cox J.C., Ingersoll J.E. and Ross S.A. 1985 A theory of the term structure of interest rates {\em Econometrica} 53, 385-407

\bibitem{Davis} Davis, M.H.A. 2004 Complete-market models of stochastic volatility, {\em Proc. Roy. Soc. Lond. A} 460, 11-26

\bibitem{HW} Hull J. and White A. 1987 The pricing of options with stochastic volatilities {\em J. Finance} 42, 281-300

\bibitem{Hull} Hull J. and White A. 1988 An analysis of the bias in option pricing caused by a stochastic volatility {\em Adv. Futures Opt. Res.} 3, 29-61

\bibitem{Scott} Scott, L. 1987 Option pricing when the variance changes randomly: theory, estimation and an application {\em J. Financial and Quantitative Analysis} 22, 419-438

\bibitem{Nel} Nelson D.B. 1990 ARCH models as diffusion approximations {\em  J. Econometrics} 45, 7-38

\bibitem{AB} Anderson T.G. and Bollerslev T. 1998 Answering the Sceptics: yes, standard volatility models do provide accurate forcasts {\em International Economic Review} 39, 885-905

\bibitem{DGH} Hobson D.G. and  Rogers L.C.G. 1998 Complete models with stochastic volatility {\em  Mathematical Finance} 8, 27-48

\bibitem{DK} Duffie D. Kan R. 1996 A yield-factor model of interest rates {\em  Mathematical Finance} 6, 379-406

\bibitem{Chern} Chernov M., Gallant A.R., Ghysels E. and Tauchen G. 2003 Alternative models for stock price dynamics {\em J. Econometrics} 116, 225-257

\bibitem{ARCH} Engle R.F. 1982 Autoregressive conditional heteroscedasticity with estimates of the variance of United Kingdom inflation {\em Econometrica} 50, 987-1007

\bibitem{KP} Kloeden P.E. and Platen E. 1992 {\em Numerical Solution of Stochastic Differential Equations} (Berlin: Springer)

\bibitem{Hair} Hairer E., Norsett S.P. and Wanner G. 1993 {\em Solving Ordinary Differential Equations} (Berlin: Springer-Verlag)

\bibitem{MAP} See http://www.math.uni-frankfurt.de/$\ensuremath{\sim}$numerik/maplestoch/

\bibitem{Gai} Gaines J.G. 1997 Variable step size control in the numerical solution of stochastic differential equations {\em SIAM J. Appl. Math.} 57, 1455-1484

\bibitem{Lam} Lamba H. 2003 An adaptive timestepping algorithm for stochastic differential equations {\em J. Comput. Appl. Math.} 161, 417-430

\bibitem{Wilk} Wilkie J. 2004  Numerical methods for stochastic differential equations {\em Phys. Rev. E} 70, 017701

\bibitem{WC} Wilkie J. and \c{C}etinba\c{s} M. 2005 Variable-stepsize Runge-Kutta methods for stochastic Schr\"{o}dinger equations {\em  Phys. Lett. A} 337, 166-182

\bibitem{ANISE} ANISE$^{\copyright}$ (available as a free trial), from Innovative Stochastic Algorithms

\bibitem{Vas} Vasicek, O. 1977 An equilibrium characterization of the term structure {\em J. Financial Economics} 5, 177-188

\end{thebibliography}
\end{document}